\documentclass[a4paper,11pt]{article}
\usepackage{jheppub} 
\usepackage{lineno}
\usepackage{amsmath}
\usepackage{comment}
\usepackage{graphicx}
\usepackage{dcolumn}
\usepackage{bm}
\usepackage{slashed}
\usepackage{braket}
\usepackage{xcolor}
\usepackage{amsfonts}
\usepackage{tikz}
\usepackage{caption}
\usepackage{subcaption}
\usepackage[hang,flushmargin]{footmisc} 
\title{\boldmath Ramp from Replica Trick}

\author{Xuchen Cao, Thomas Faulkner}
\affiliation{Department of Physics, University of Illinois,
1110 W. Green St., Urbana, IL 61801-3080, USA}

\emailAdd{xuchenc2@illinois.edu}

\abstract{
We compute the spectral form factor of the modular Hamiltonian  $K=-\ln\rho_A$ associated to the reduced density matrix of a Haar random state. A ramp is demonstrated and we find an analytic expression for its slope. Our method involves an application of the replica trick, where we first calculate the correlator $\braket{\text{tr}\rho_A^n\;\text{tr}\rho_A^m}$ 
at large bond dimension and then analytically continue the indices $n,m$ from integers to arbitrary complex numbers. We use steepest descent methods at large modular times to extract the ramp. The large bond dimension limit of the replicated partition function is dominated by a sum over \emph{annular non-crossing permutations}. We explored the similarity between our results and calculations of the spectral form factor in low dimensional gravitational theories where the ramp is determined by the double trumpet geometry. We find there is an underlying resemblance in the two calculations, when we interpret the annular non-crossing permutations as representing a discretized version of the double trumpet. Similar results are found for an equilibrated pure state in place of the Haar random state.
}

\begin{document}
\maketitle
\flushbottom

\section{Introduction}
\label{sec:intro}
\begin{figure}
	\centering
		\includegraphics[width=0.6\textwidth]{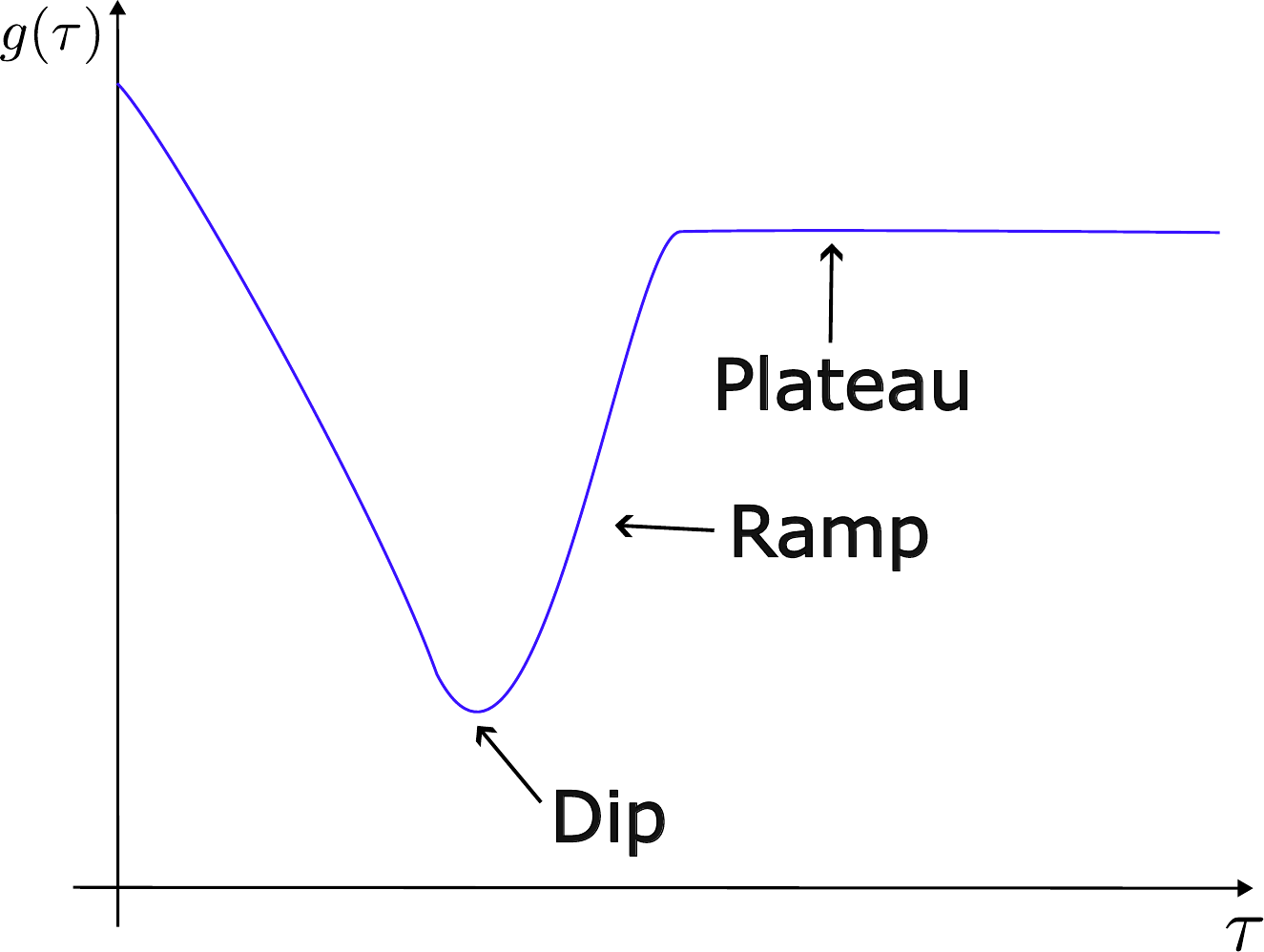}
        \caption{A sketch of the dip-ramp-plateau behaviour of the spectral form factor, $g(\tau)$ decays for small $\tau$, but at some \emph{dip time} $\tau=\tau_d$, $g(\tau)$ stops decreasing but instead enters a region of strict linear growth which is usually called the \emph{ramp}. Finally $g(\tau)$ saturates at some $\tau=\tau_p$ and then remains a constant, this region is called the \emph{plateau}. The slope of the ramp and the height of the plateau had been greatly exaggerated, the plateau is in fact suppressed by a factor $e^{-S}$ where $S$ is the entropy of the system.}
        \captionsetup{format=hang}
        \label{dip_ramp_plateau}
\end{figure}

The study of the development of chaos in quantum systems has a long history (see, for example~\cite{haake1991quantum,d2016quantum} for reviews). Quantum chaos is related to many interesting topics such as thermalization~\cite{deutsch1991quantum,srednicki1994chaos}, information 
scrambling~\cite{maldacena2016bound,sekino2008fast,fisher2023random,mi2021information}, and blackholes~\cite{sekino2008fast,shenker2014multiple,shenker2014black,magan2018black}. 
There are various ways to characterize quantum chaos including out-of-time-order correlation functions and related quantities~\cite{larkin1969quasiclassical,hashimoto2017out,maldacena2016bound,shenker2015stringy,peres1984stability,jalabert2001environment}, which characterize chaos at early times, as well as
level repulsion in the energy spectrum ~\cite{bohigas1984characterization,von1993merkwurdige,wigner1958distribution,wigner1993characteristic1,wigner1993characteristic2}, which pertains to chaos on much longer time scales. 
In particular, Bohigas, Gianonni and Schmidt~\cite{bohigas1984characterization}
proposed that the spectrum of a quantum chaotic system always exhibits universal features which is the same as that of a random Hamiltonian matrix ensemble in the appropriate symmetry class. Therefore many properties of quantum chaotic systems can be deduced from the theory of random matrices~\cite{mehta2004random,eynard2015random}.\\ \\
In characterizing the spectral properties of a Hamiltonian matrix, the following quantity called \emph{spectral form factor} is widely used~\cite{cotler2017black}:
\begin{equation}\label{spectral form factor}
    g(\tau)=\braket{\sum_{i,j} e^{-i\tau(E_i-E_j)}}
\end{equation}
where $E_i,E_j$ are eigenvalues of the Hamiltonian matrix $H$, and $\braket{\ldots}$ is some ensemble average which removes non-universal rapid oscillations in $g(\tau)$. At $\tau=0$ $g(\tau)$ is of order $e^{2S}$ where $S$ is the entropy of the system. In the case of trivial correlation between $E_i$'s, $g(\tau)$ decays to order $e^{S}$ as some negative power of $\tau$. However, in the presence of non-trivial spectral correlation the spectral form factor takes a well known dip-ramp-plateau form as summarized in Figure~\ref{dip_ramp_plateau}. 
Thus the ramp and the plateau serves as a diagnostic of non-trivial spectral correlation, and therefore quantum chaos.
In some cases, $g(\tau)$ can be analytically obtained from certain random Hamiltonian matrix ensembles, examples include the GUE/GOE ensemble~\cite{cotler2017black} and the Wishart ensemble~\cite{chen2018universal}. Similar results can be found for the SYK model~\cite{cotler2017black,saad2018semiclassical}. For theories with a holographic dual one can study the ramp and plateau from a bulk perspective~\cite{saad2018semiclassical,saad2019jt} where in particular the ramp can be obtained from the gravitational path integral with the boundary condition of two thermal circles. \\ \\
In the present paper, we are interested in the spectral correlation of a specific Hamiltonian called the modular(or entanglement) Hamiltonian, which is defined in the following way: take a bipartite system $A\cup\bar A$ and a state $\ket{\psi}\in\mathcal H_{A\cup \bar A}$, then the modular Hamiltonian is 
\begin{equation}
    K=-\ln(\rho_A)
\end{equation}
where $\rho_A=\text{tr}_{\bar A} \ket{\psi}\bra{\psi}$. There are a few different reasons to consider the modular Hamiltonian. Firstly, if we adopt a strong version of the eigenstate thermolization hypothesis (ETH) as proposed in~\cite{garrison2018does}, the modular Hamiltonian $K_A$ for some subregion $A$ obtained by partially tracing a typical state in a thermalizing system actually encodes information about the projection of the physical Hamiltonian $H$ onto the subsystem $A$, at least if we ignore the correlation between $A$ and $\bar A$ across the boundary. Secondly, the modular Hamiltonian represents a vast generalization of the physical Hamiltonian of the system allowing for a potentially more local characterization of quantum chaos, similar to the generalization from a boundary thermal state which is dual to a bulk black hole with its entropy determined by the Bekenstein-Hawking entropy-area relation to a subregion on the boundary which is dual to a bulk entanglement wedge with its entanglement entropy determined by the Ryu-Takayanagi formula~\cite{ryu2006holographic}. For example in holographic systems, to the leading order in $\frac{1}{G_N}$, the bulk dual of the boundary modular Hamiltonian is the area operator of the corresponding extremal surface~\cite{jafferis2016gravity,jafferis2016relative}.\\ \\
In~\cite{chen2018universal} it was already demonstrated numerically that for some specific systems $K$ indeed exhibits a ramp. In this paper we will study the entanglement Hamiltonian $\rho_A$ obtained by partially tracing a Haar random state and give an analytical proof for the existence of the ramp using the replica trick. Our strategy is to first calculate $\braket{\text{tr}(\rho_A^n)\text{tr}(\rho_A^{m})}$ for integer $n,m$, where $\braket{\ldots}$ denotes an average over $\ket{\psi}$ which is chosen from a Haar ensemble. Then we relate the result to combinatorial mathematical objects called \emph{annular non-crossing permutation}. Such permutations arise in the study of second order asymptotics of random matrices \cite{mingo2004annular} and have also arisen in the study of reflected entropy in random tensor networks \cite{akers2022reflected,akers2023reflected,akers2024reflected}.
Finally we implement the analytical continuation $n,m\rightarrow \pm is$, where $s$ is the real time, to identify the linear ramp. One of the advantages to use the replica trick is that it gives a well defined boundary condition in the holographic duality setup which naturally leads to a bulk dual. So in principle this formalism allows us to find a bulk geometric interpretation of the ramp of an entanglement Hamiltonian which does not depend on the time translation symmetry on the boundary, as in the cases discussed in~\cite{saad2018semiclassical,saad2019jt}. Our result includes an expression for the slope of the ramp for arbitrary temperature as well as subsystem size, which is especially simple for an equally bipartite system, in this case the ramp takes the form
\begin{equation}
    g^{con}(\beta,\tau)\sim\frac{4^{2\beta}}{4\pi N_A^{2\beta}}\cdot \frac{\tau}{\beta}
\end{equation}
where $\beta$ is the inverse temperature, $g^{con}(\beta,\tau)$ is the connected part of $g(\beta,\tau)$, which is responsible for the linear ramp, and the numerical factor comes from the normalization of the density matrix. Note that in this case the scaling form $\tau/\beta$ agrees with the result in~\cite{saad2019jt} obtained from JT gravity calculations.\\ \\
The remainder of this paper is organized as follows, in Section~\ref{sec:random} we introduce the relationship between random matrix theory and annular non-crossing permutations and compute the correlator $\braket{\text{tr}(\rho^{n})\text{tr}(\rho^m)}$ for non-negative integers $n,m$. In Section~\ref{sec:diagrammatic} we give physical interpretations of annular non-crossing permutations from different physical perspectives. In Section~\ref{sec:analytical} we analytically continue the correlator to find an expression for the spectral form factor and prove the existence of a linear ramp. In Section~\ref{sec:geodesic} we view annular non-crossing permutations from a new perspective which indicates a possible relationship to ramp calculations in JT gravity. Finally in Section~\ref{sec:conclusion} we conclude the article and propose some future directions.

\section{Random Matrix and Annular Non-Crossing Permutation}
\label{sec:random}
We start from a bipartite system with a Hilbert space $\mathcal H=\mathcal H_A\otimes \mathcal H_B$, where $A$ and $B$ are the two subsystems with Hilbert space dimensions $\dim(H_{A,B})=N_{A,B}$. Let $\ket{\psi}\in\mathcal H$, which can be written as
\begin{equation}
    \ket{\psi}=\sum_{iJ}X_{iJ}\ket{i_A}\otimes\ket{J_B}
\end{equation}
Here $i=1,2,\ldots N_A$ and $J=1,2,\ldots N_B$. The reduced density matrix for subsystem A is thus
\begin{equation}
    \rho_A=XX^\dagger
\end{equation}
Now we assume that the state $\ket{\psi}$ is picked from the Haar ensemble, that is, $\ket{\psi}=U\ket{\psi_0}$, where $\ket{\psi_0}$ is a fiducial state and $U$ is a random unitary matrix. For the Haar ensemble the measure of $U$ satisfies $dU=d(VU)=d(UV)$ for arbitrary unitary matrix $V$. A useful way to visualize this is to think of the coefficients $X_{iJ}$ as a vector in the $2N_AN_B$ dimensional Euclidean space (since it has $2N_AN_B$ real components) with unit norm, which comes from the normalization. The Haar ensemble then corresponds to a random sampling of the vector on the sphere of unit norm with constant probability distribution. When $N_{A,B}$ are large, we can ignore the unit norm constraint and treat each component of $X_{iJ}$ as independent~\cite{chen2018universal}, this comes from the fact that the volume of a high dimensional ball concentrates around its surface. For later convenience we will take each $X_{iJ}$ as a complex random variable with Gaussian distribution $\mathcal N(0,1)$. With this normalization we have
\begin{equation}
    \braket{\text{tr}\rho_A}=N_AN_B
\end{equation}
and thus if we neglect the fluctuation of the trace we have
\begin{equation}
    \rho_A=\frac{XX^\dagger}{N_AN_B}
\end{equation}
This is known as the Wishart random matrix ensemble. In~\cite{mingo2004annular} it had been shown that correlation functions involving $\rho_A$ can be related to combinatorial mathematical objects called annular non-crossing permutations $\text{ANC}(n,m)$, which are a class of permutations in the permutation group $S_{n+m}$. See Appendix~\ref{append:annular} for a brief review of annular non-crossing permutations. The result we will make use of is 
\begin{equation}\label{correlator}
     \braket{\text{tr}\rho_A^n\;\text{tr}\rho_A^m}_c=\frac{1}{N_A^{n+m}}
     \sum_{\substack{\tau\in ANC\\ \tau\;\text{connected}}}\lambda^{\#(\tau)}+O(\frac{1}{N_A^{n+m+1}})
\end{equation}
here we take the limit where $N_A,N_B\rightarrow \infty$ while keeping the ratio $\lambda=\frac{N_A}{N_B}$ fixed. The subscript $c$ stands for the connected part of the correlator, which gives the linear ramp after analytical continuation. The sum is over all connected annular non-crossing permutations $\tau\in\text{ANC}(n,m)$, with $\#(\tau)$ the number of orbits in $\tau$. Without loss of generality, we can assume that $\lambda\leq 1$, this is due to a constraint which states that an arbitrary connected annular non-crossing permutation $\tau$ satisfies the following identity 
\begin{equation}\label{geodesic condition}
    \#(\tau)+\#(\tau^{-1}\circ \gamma_0)=n+m
\end{equation}
where $\gamma_0=(12\ldots n)(n+1,n+2\ldots n+m)$. Thus if $N_B<N_A$ we can reorganize the sum into  
\begin{equation}\label{correlator2}
     \braket{\text{tr}\rho_A^n\;\text{tr}\rho_A^m}_c=\frac{1}{N_B^{n+m}}
     \sum_{\substack{\tilde\tau\in ANC\\ \tilde\tau\;\text{connected}}}\tilde\lambda^{\#(\tilde\tau)}+O(\frac{1}{N_B^{n+m+1}})
\end{equation}
where $\tilde\lambda=\frac{1}{\lambda}$ and $\tilde \tau= \tau^{-1}\circ \gamma_0$ (which is also connected and annular non-crossing). This sum takes the same form as in the $\lambda\leq 1$ case except for a different overall factor. From now on we will usually assume that $\lambda\leq 1$ for simplicity.\\ \\
Now we evaluate Equation~\ref{correlator}. To do the sum we need to enumerate connected annular non-crossing permutations with a given number of orbits, $\#(\tau)$. This had been studied in~\cite{kim2013cyclic} where it was shown that the number of all annular non-crossing permutations with $c$ connected orbits, $r$ exterior orbits and $s$ interior orbits is given by
\begin{equation}
      \#\text{ANC}(n,m;c,r,s)=c\binom{n}{r}\binom{m}{s}\binom{n}{r+c}\binom{m}{s+c}
\end{equation}
We can split the sum over $\tau$ into sums over $c,r,s$. The sums over $r,s$ can be done explicitly in terms of hypergeometric functions~\cite{hayden1986summation}, which gives
\begin{equation}\label{sum1}
    \sum_{\substack{\tau\in ANC\\ \tau\;\text{connected}}}\lambda^{\#(\tau)}=\sum_{c=1}^{\infty}c\lambda^c\binom{m}{c}\binom{n}{c}
    {}_2F_1(\substack{-n,-n+c\\ c+1}|\lambda){}_2F_1(\substack{-m,-m+c\\ c+1}|\lambda)
\end{equation}
Note that the binomial coefficients in fact cut the sum at $c=\min(n,m)$. Equation~\ref{sum1} is one of the key results of this article, in the following we will do the analytical continuation and show that this leads to a linear ramp. Before doing that, however, we will first give some physical reasoning how annular non-crossing permutations enter our calculation.

\section{Physical Origin of Annular Non-Crossing Permutations}
\label{sec:diagrammatic}
In this section we explore the origin of annular non-crossing permutations in our calculation of $\braket{\text{tr}\rho_A^n\;\text{tr}\rho_A^m}_c$ from a few different perspectives. The first approach is a generalization of the equilibrium formalism introduced in~\cite{liu2021entanglement} to multi-trace correlators, which enables us to expand our discussion to more general cases beyond the Haar ensemble, such as the micro-canonical ensemble. Then we will view the problem from a holographic perspective, that is, we assume a bulk dual of our theory and explain what bulk geometries gives rise to annular non-crossing permutations. 

\subsection*{The Equilibrium Approach}
To start, we review the equilibrium formalism proposed in~\cite{liu2021entanglement} and introduce some notations, then we generalize to the calculation of our double trace correlators. The annular non-crossing permutations will emerge naturally from this perspective. Similar structures can also be found in~\cite{nakagawa2018universality,fujita2018page}.\\ \\ 
Suppose that we start from a fiducial state $\ket{\Psi_0}$, and apply the evolution operator $U(t)$, after some time $t>t_s$ the density matrix of the system becomes $\rho(t)=U\ket{\Psi_0}\bra{\Psi_0}U^\dagger$, depending on the details of our model $\rho(t)$ can be approximated by an \emph{equilibrium density matrix} $\rho_{eq}$ when we calculate certain observables. Furthermore, $\rho_{eq}$ is supposed to be invariant under time evolution, that is $U\rho_{eq}U^\dagger=\rho_{eq}$. Some examples include
\begin{itemize}
    \item For a system in which all states can be accessed, the evolution can be taken to be Haar random while the final density matrix can be taken as identity, this is the case we will focus on for most of the time
    \item For a fiducial state which includes energy eigenstates in a narrow energy window, the final density matrix can be taken as a micro-canonical one
    \item For a fiducial state which includes a broad range of energy eigenstates, the final density matrix can be taken as a canonical one at some fixed temperature. 
\end{itemize}
Let our bipartite system be $A\cup B$ with $N_{A/B}$ the Hilbert space dimension of the two subsystems. The Renyi entropy of the reduced density matrix $\rho_A$ is given by
\begin{equation}\label{latetimetrace}
    S_A^{(n)}=\text{tr}\rho_A^n=\text{tr}_A(\text{tr}_B U\ket{\Psi_0}\bra{\Psi_0}U^\dagger)^n
\end{equation}
And similarly the double trace correlator we want to study can be written as
\begin{equation}\label{latetimedoubletrace}
    \text{tr}\rho_A^n\;\text{tr}\rho_A^m=\text{tr}_A(\text{tr}_B U\ket{\Psi_0}\bra{\Psi_0}U^\dagger)^n\text{tr}_A(\text{tr}_B U\ket{\Psi_0}\bra{\Psi_0}U^\dagger)^m
\end{equation}
Here we have not included the average since at this stage we have not introduced any ensemble average. In this subsection we use $\braket{\ldots}$ for results obtained from the equilibrium approximation, which amounts to replace the late-time state with a density matrix drawn from suitable ensembles. Now let $\ket{ij}$ denote an orthonormal basis for our Hilbert space, with $i$ and $j$ for subsystems A and B respectively. And we define the following \emph{effective identity operator}
\begin{equation}
    \rho_{eq}=\frac{I_\alpha}{Z(\alpha)}
\end{equation}
where $\alpha$ denotes the ensemble we choose to approximate the late-time state, as discussed above, $\rho_{eq}$ and $Z(\alpha)$ are the respective density matrix and partition function. For example in the canonical ensemble case $I_\alpha=e^{-\beta H}$ and $Z(\alpha)$ is just the canonical partition function. Furthermore we define 
\begin{equation}
    Z_n(\alpha)=\text{tr} I_\alpha^n
\end{equation}
After applying the equilibrium approximation (see Appendix~\ref{append:equi} for details). We obtain the following result for the double trace correlator
\begin{equation}\label{double trace equilibrium approximation}
    \braket{\text{tr}\rho_A^n\;\text{tr}\rho_A^m}_c=\frac{1}{Z_1^{m+n}}\sum_{\substack{\{i\},\{j\}\\ \tau\;\text{connected}}}
    \braket{i_{\tilde\tau(1)}j_{\tau(1)}|I_\alpha|i_1j_1}\ldots \braket{i_{\tilde\tau(n+m)}j_{\tau(n+m)}|I_\alpha|i_{n+m}j_{n+m}}
\end{equation}
Here $\tau\in S_{n+m}$, $\tilde\tau=\gamma_0^{-1}\cdot \tau$. As we are interested in the connected part of the correlator we have restricted the sum to be over connected permutations only. In the Haar random case where $I_\alpha=\mathbb I$, it is clear that each term in the sum gives a factor $N_A^{\#(\tilde\tau)}N_B^{\#(\tau)}$. We assume that the same scaling behaviour holds even in the $I_\alpha\neq \mathbb I$ case as each trace gives a large factor of order $N_{A,B}$. Since $N_A$ and $N_B$ are always taken to be of the same order, the leading contribution to the sum comes from terms which maximize $\#(\tilde \tau)+\#(\tau)$. Equation~\ref{geodesic condition} tells that for connected permutations we have $\#(\tilde \tau)+\#(\tau)\leq m+n$, where the equality holds when $\tau$ is an annular non-crossing permutation. Therefore, we see that annular non-crossing permutations naturally arise in our calculation.

\subsection*{Micro-Canonical Ensemble}
Now we apply the formalism discussed above to micro-canonical ensembles, for which we will obtain a result for $\braket{\text{tr}\rho_A^n\;\text{tr}\rho_A^m}_c$ whose form is a simple generalization of the Haar random case. This form allows us to identify a ramp in its spectral form factor, following the derivation in Section~\ref{sec:analytical}. In the microscopic canonical ensemble, the equilibrium density matrix $\rho_{eq}$ is 
\begin{equation}
    \rho_{eq}=\begin{cases}
        \frac{1}{N_I}\;\;(E\in I)\\
        0\;\;(\text{others})
    \end{cases}
\end{equation}
Here $I$ is a narrow energy window around some energy $E_I$, $N_I$ is the total number of states within this energy window. From now on we will assume that the total system is a weakly interacting one which satisfies $E\sim E_A+E_B$. This condition is usually satisfied in the thermodynamic limit at least when the interaction is short range. In this case we substitute~\cite{liu2021entanglement}
\begin{equation}\label{micro projector}
    I_\alpha=\sum_{E_A+E_B\in I} P_A(E_A) P_B(E_B)
\end{equation}
here $P_A(E_A)$ is a projector in the subspace $\mathcal H_A$ projecting onto states with energy $E_A$, similarly we define $P_B(E_B)$. Plugging this form of $I_\alpha$ into Equation~\ref{double trace equilibrium approximation} we get (See Appendix~\ref{append:micro} for details)
\begin{equation}\label{microcanonical double trace}
    \braket{\text{tr}\rho_A^n\;\text{tr}\rho_A^m}_{c,micro}=\frac{1}{N_I^{n+m}}
    \sum_{E_A} \sum_{\tau\in \text{ANC}(n,m)} (d^A_{E_A})^{\#(\tilde\tau)}(d^B_{E_I-E_A})^{\#(\tau)}
\end{equation}
Here we have defined $d^A_{E_A}$ to be the number of states in the subspace $\mathcal H_A$ with energy $E_A$, and the same for $d^B_{E_B}$. We again sum over connected annular non-crossing permutations to obtain the leading order contribution to the connected part. Here we have obtained for each $E_A$ a sum of the form~\ref{sum1}, with a final sum over $E_A$. The derivation in Section~\ref{sec:analytical} is applicable to each $E_A$ sector, providing a non-trivial contribution to the ramp, and the total amplitude of the ramp can be calculated in principle by completing the sum over $E_A$, which depends on the details of the energy level structure of the system.

\subsection*{A Bulk Interpretation}
\begin{figure}
	\centering
		\includegraphics[width=0.5\textwidth]{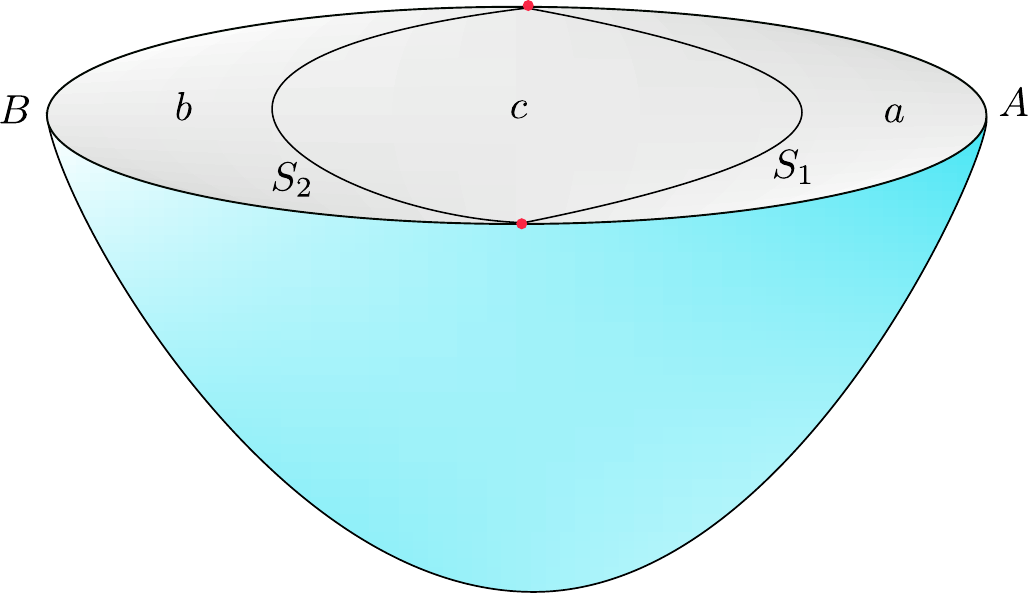}
        \captionsetup{format=hang}
        \caption{A fixed area state with area $A_{1,2}$ for diffeomorphism invariantly defined surfaces $S_{1,2}$, prepared by an Euclidean path integral in the bulk with certain boundary conditions}
        \label{fixed area state}
\end{figure}

\begin{figure}
	\centering
		\includegraphics[width=0.8\textwidth]{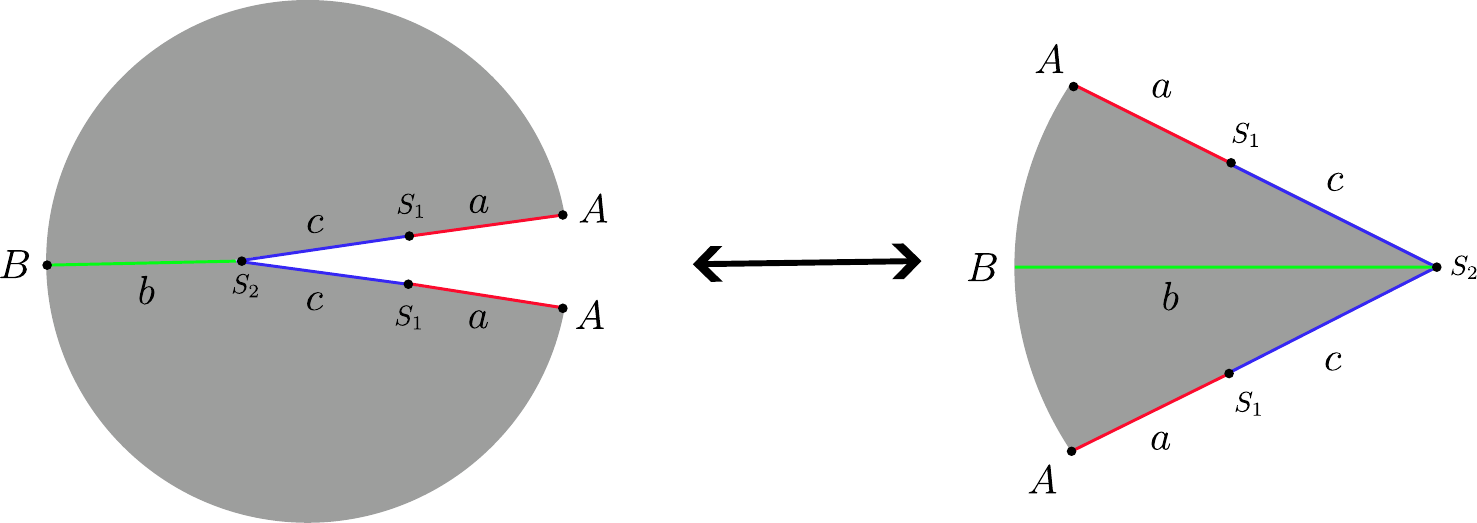}
        \captionsetup{format=hang}
        \caption{Gluing two copies of the state $\ket{\psi}$ along boundary subregion $B$ and bulk subregion $b$ to produce the reduced density matrix $\rho_A$}
        \label{gluing density matrix}
\end{figure}

\begin{figure}
	\centering
	\begin{subfigure}{.7\textwidth}
		\includegraphics[width=\textwidth]{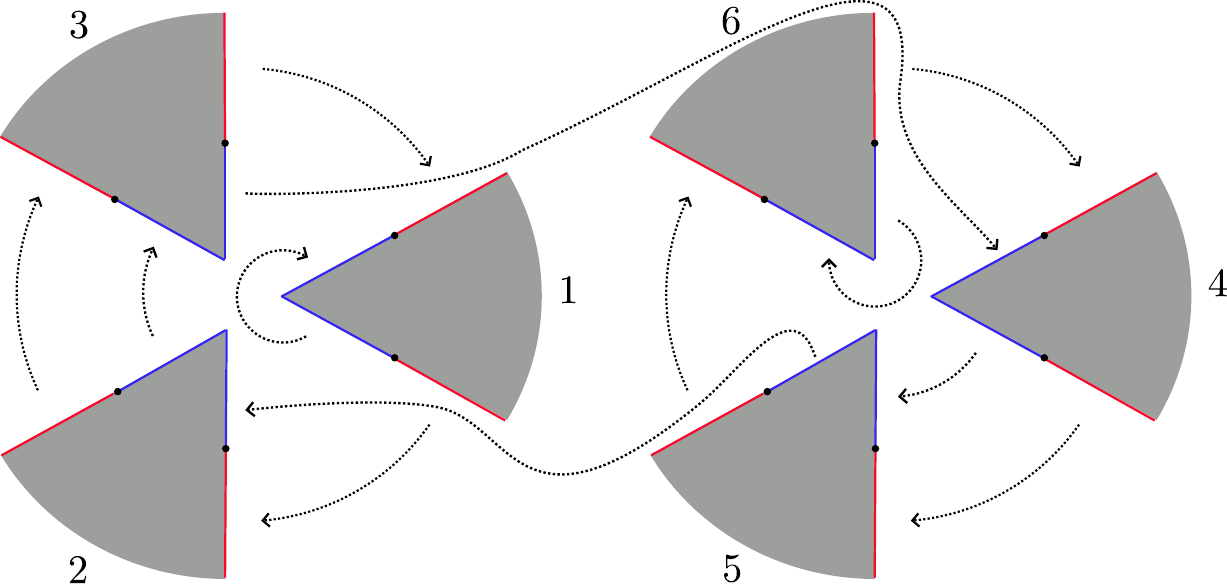}
        \captionsetup{format=hang}
		\caption{A bulk saddle for the double trace with $\tau=(1)(2,3,4,5)(6)$}
        \label{double trace 1}
        \vspace*{0.7cm}
	\end{subfigure}
	\begin{subfigure}{.7\textwidth}
		\includegraphics[width=\textwidth]{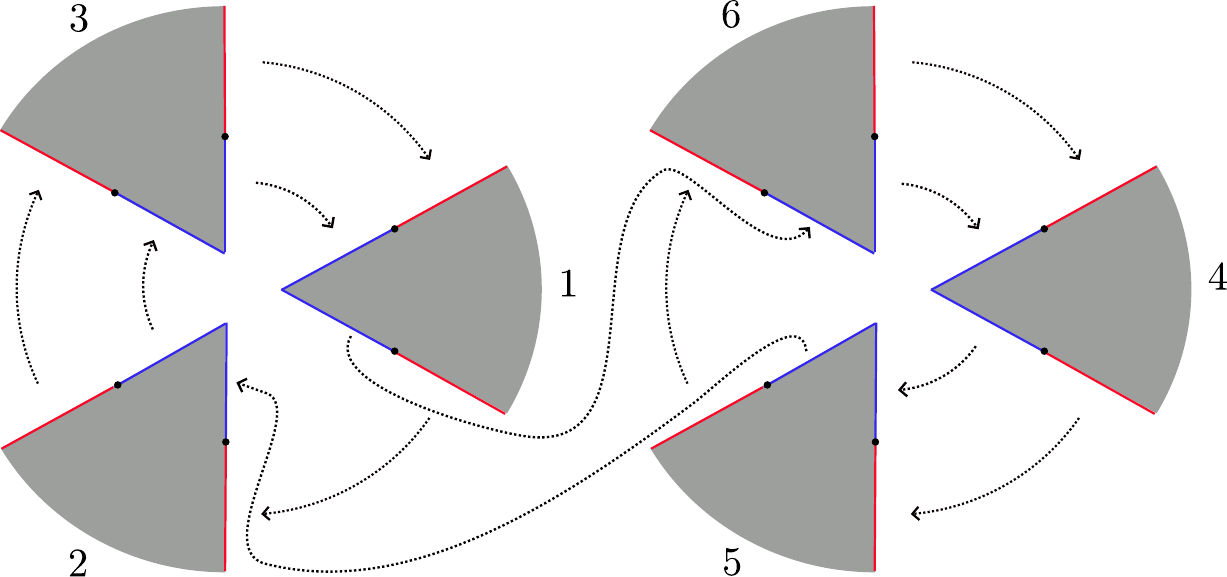}
        \captionsetup{format=hang}
        \caption{A bulk saddle for the double trace with $\tau=(1,2,3,6,4,5)$}
	\label{double trace 2}
        \end{subfigure}
        \captionsetup{format=hang}
        \caption{Examples of bulk saddles for the double trace $\braket{\text{tr}\rho_A^n\;\text{tr}\rho_A^m}$ with $n=m=3$} 
        \label{bulk saddles of double trace}
\end{figure}

The derivation above is based on statistical mechanics and does not assume the theory to be holographic. Now we discuss how annular non-crossing permutations emerge from a bulk perspective in the presence of a bulk dual, this will be done in a way parallel to the calculation of the Renyi entropy in the presence of two competing RT surfaces in~\cite{akers2021leading}. \\ \\
We work with the so called fixed area states~\cite{dong2019flat}, these are bulk states in which some diffeomorphism invariant surfaces have fixed areas. Again we assume that the boundary is bipartite and denote the two subregions with $A$ and $B$. To start, we construct a bulk state $\ket\psi$ with two different local minimal surfaces $S_1$ and $S_2$ anchored on the boundary entanglement surface $A\cap B$, and with areas $A_1$ and $A_2$, which are of the same order. The bulk is then divided into three regions: $a$ enclosed by $A$ and $S_1$, $b$ enclosed by $B$ and $S_2$, and $c$ enclosed by $S_1$ and $S_2$, see Figure~\ref{fixed area state}.
A state prepared by a path integral with these boundary conditions has analogous entanglement structure to that of a typical state in the Haar ensemble where
the ratio $N_A/N_B$ is determined by the ratio of areas:
\begin{equation}\label{area definition}
\frac{N_A}{N_B} = e^{(A_1-A_2)/4G}
\end{equation}
In particular a typical state has an entanglement entropy with small fluctuation in the large $N$ limit as predicted in~\cite{page1993average} and the competition between the two RT surfaces corresponds to the Page transition at $N_A/N_B=1$. In order to prepare $\ket{\psi}$ we need to do the Euclidean path integral as shown in Figure~\ref{fixed area state}, here we assumed time reflection symmetry for simplicity. To get the reduced density matrix $\rho_A$ we need to take two copies of the Euclidean path integral and glue them along the boundary subregion $B$ as well as the bulk subregion $b$, this gluing operation is schematically depicted in Figure~\ref{gluing density matrix}.
Next we do the replica trick to obtain $\text{tr}\rho_A^n$, this is done by gluing boundary subregion $A$'s, as well as the bulk subregion $a$'s in a cyclic way. However, there remains the possibility that bulk subregion $c$'s in each copy are not glued together in a cyclic way, in fact, they can be glued in any order specified by an arbitrary permutation. With this setup we can interpret our double trace correlator $\braket{\text{tr}\rho_A^n\;\text{tr}\rho_A^m}$ in the following way: we take $n+m$ copies of $\rho_A$ and glue boundary subregion $A$'s and bulk subregion $a$'s in a cyclic manner for the first $n$ and the remaining $m$ copies, while the $c$'s are glued together using an arbitrary permutation $\tau$, some examples are shown in Figure~\ref{bulk saddles of double trace}.
Since we are interested in the connected part of the correlator we take $\tau$ to be connected. Now we calculate the partition function corresponding to such a bulk configuration. Assuming semiclassical bulk we have
\begin{equation}
    \braket{\text{tr}\rho_A^n\;\text{tr}\rho_A^m}_c=\sum_{\tau\;\text{connected}} \frac{e^{-I_{n,m}(\tau)}}{e^{-(n+m)I_1}}
\end{equation}
here $I_{n,m}(\tau)$ is the bulk action of a configuration described by $\tau$ and $I_1$ is the bulk action obtained from calculating $\braket{\psi|\psi}$ holographically. The denominator comes from normalization $\text{tr}\rho_A=1$. And the semiclassical limit implies that the equation of motion for the metric is satisfied everywhere in the bulk, except at RT surfaces which are singular.\\ \\
Now what remains is to calculate the action $I_{n,m}$ as well as $I_1$. The insertion of fixed area surfaces results in conical singularities, and we denote the opening angle around $S_1$ and $S_2$ with $\phi_1$ and $\phi_2$ respectively\footnote{In principle, a fixed area surface does not have a well defined opening angle, however at the level of bulk saddle point analysis we can neglect the uncertainty of opening angles and take them to be definite values, see discussion in~\cite{dong2019flat}}. The replicated geometry agrees with the $n=1$ geometry away from these conical singularities, therefore the only difference between the denominator and the numerator comes from singularities. It is well known that for a conical singularity with deficit angle $(\theta-2\pi)$ there is a contribution to the gravitational action due to the divergence of Ricci tensor $R_{\mu\nu}$ on the singular hypersurface, which is given by
\begin{equation}
    I_{conical}=\frac{(\theta-2\pi)A}{8\pi G}
\end{equation}
Here $A$ is the area of the singular surface. Therefore we can obtain $I_{n,m}$ by counting conical singularities as well as their deficit angles, and adding up their contributions, the result is
\begin{equation}
    I_{n,m}(\tau)=\frac{\big[(n+m)\phi_1-\#(\tau)\cdot 2\pi\big] A_1}{8\pi G}+\frac{\big[(n+m)\phi_1-\#(\tau^{-1}\circ \gamma_0)\cdot 2\pi\big] A_2}{8\pi G}
\end{equation}
while the singularity contribution to the denominator is 
\begin{equation}
    (n+m)I_1=(n+m)\frac{(\phi_1-2\pi)A_1+(\phi_2-2\pi)A_2}{8\pi G}
\end{equation}
Combining the results above we get
\begin{equation}
    \braket{\text{tr}\rho_A^n\;\text{tr}\rho_A^m}_c=
    e^{-\frac{(n+m)(A_1+A_2)}{4G}}\sum_{\tau\;\text{connected}} \exp\bigg[\frac{\#(\tau)}{4G}A_1+\frac{\#(\tau^{-1}\circ \gamma_0)}{4G}A_2\bigg]
\end{equation}
following the same reasoning as in the equilibrium approach, if $A_1$ and $A_2$ are of the same order then the leading order contribution comes from annular non-crossing permutations which satisfy Equation~\ref{geodesic condition}. Thus we obtain an expression which takes the same form as Equation~\ref{correlator}, with the identification Equation~\ref{area definition}.

\section{Analytical Continuation and the Ramp}
\label{sec:analytical}
\begin{figure}
	\centering
		\includegraphics[width=0.6\textwidth]{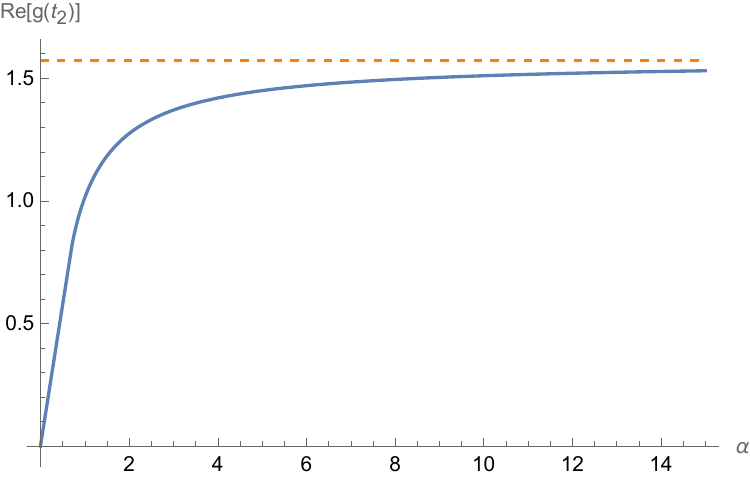}
        \captionsetup{format=hang}
	\caption{A plot of the function $\text{Re}[G(t_2)]$ at $\lambda=0.1$, which grows linearly for $\alpha<\alpha^*$, and saturates the upper bound $\frac{\pi}{2}$ as $\alpha$ becomes large} 
        \label{Regt_2}
\end{figure}

\begin{figure}
	\centering
	\begin{subfigure}{.45\textwidth}
		\includegraphics[width=\textwidth]{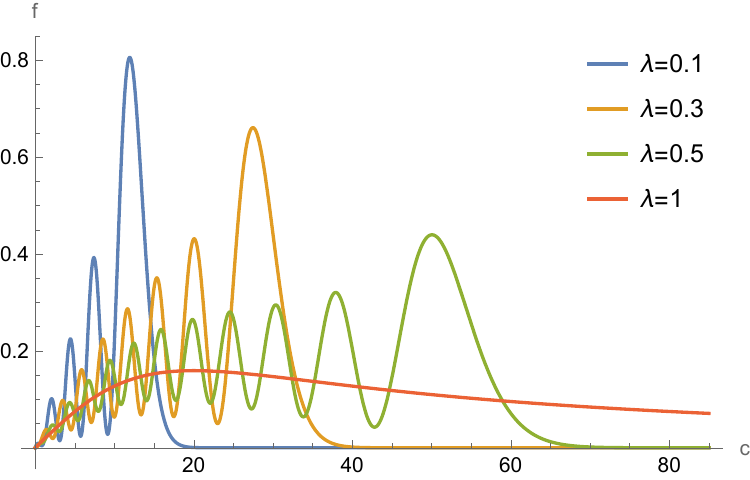} 
		\caption{$f_\beta(c,s)$ as a function of $c$ for $s=20$ and $\beta=0$}
        \label{f for beta=0}
	\end{subfigure}
        \hspace{2em}
	\begin{subfigure}{.45\textwidth}
		\includegraphics[width=\textwidth]{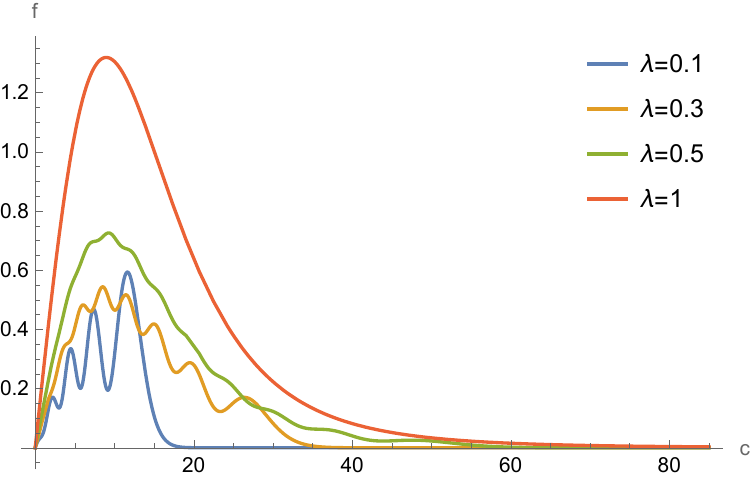}
        \captionsetup{format=hang}
        \caption{$f_\beta(c,s)$ as a function of $c$ for $s=20$ and $\beta=1$}
	\label{f for beta=1}
        \end{subfigure}
        \captionsetup{format=hang}
        \caption{Illustrations of $f_\beta(c,s)$ for different values of $\beta$ and $\lambda$. It is easy to see that non-trivial $\beta$ suppresses $f_\beta(c,s)$ for large values of $c$, and therefore cures the divergence of the slope at $\lambda=1$}
        \label{behaviour of f}
\end{figure}

\begin{figure}
	\centering
	\begin{subfigure}{.45\textwidth}
		\includegraphics[width=\textwidth]{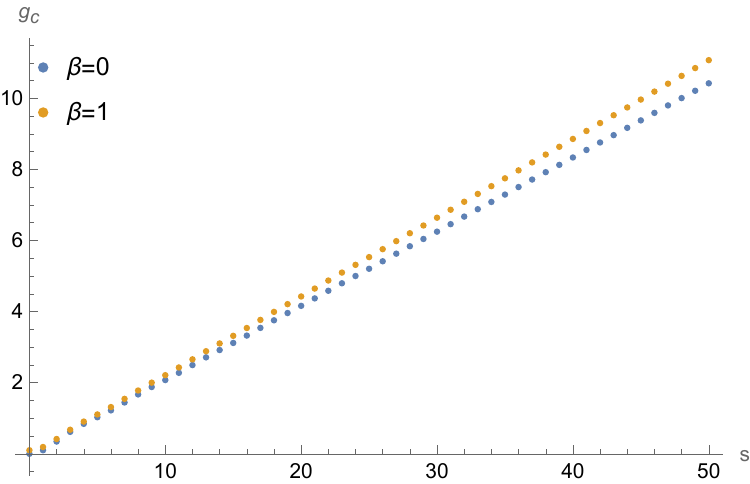} 
		\caption{The ramp for $\lambda=0.1$}
        \vspace*{0.7cm}
        \label{ramp 0.1}
	\end{subfigure}
        \hspace{2em}
	\begin{subfigure}{.45\textwidth}
		\includegraphics[width=\textwidth]{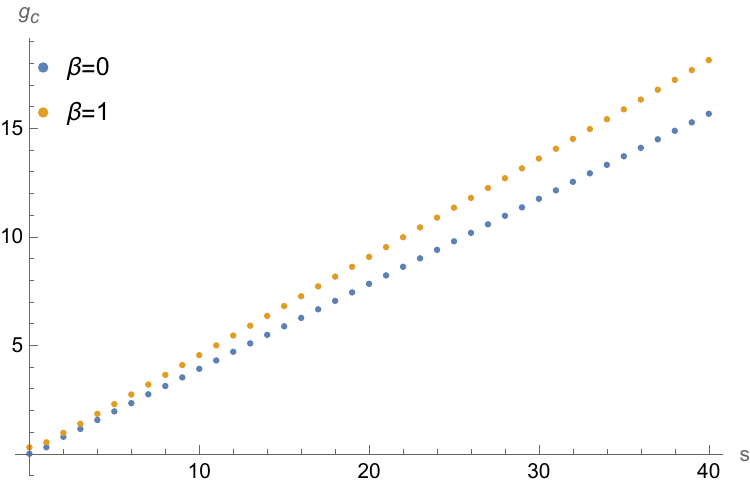}
        \captionsetup{format=hang}
        \caption{The ramp for $\lambda=0.3$}
        \vspace*{0.7cm}
	\label{ramp 0.3}
        \end{subfigure}
        \begin{subfigure}{.45\textwidth}
		\includegraphics[width=\textwidth]{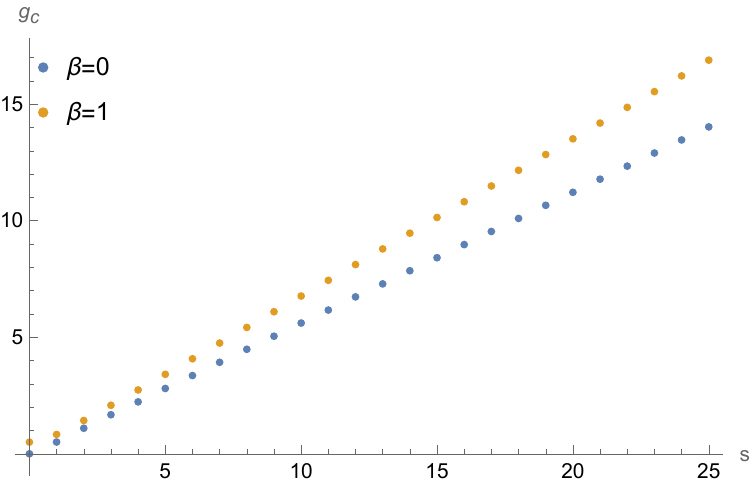}
        \captionsetup{format=hang}
        \caption{The ramp for $\lambda=0.5$}
	\label{ramp 0.5}
        \end{subfigure}
        \hspace{2em}
        \begin{subfigure}{.45\textwidth}
		\includegraphics[width=\textwidth]{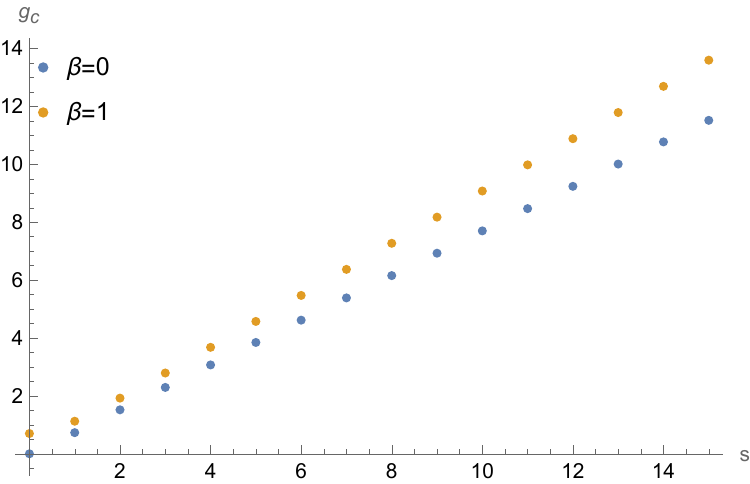}
        \captionsetup{format=hang}
        \caption{The ramp for $\lambda=0.7$}
	\label{ramp 0.7}
        \end{subfigure}
        \captionsetup{format=hang}
        \caption{Numerical results of the sum $g^{con}_\beta(s)=\sum_c f_\beta(c,s)$ for different values of $\lambda$ and $\beta$. We choose different upper bounds for $s$ for different $\lambda$'s as the number of terms to be summed over increases very rapidly as $\lambda$ increases}
        \label{numerics for ramps}
\end{figure}

In this section we analytically continue Equation~\ref{sum1} to imaginary values of its arguments and use an asymptotic argument to show that it indeed gives a linear ramp. Furthermore we will derive a formula which determines the amplitude of the ramp. The connected part of the spectral form factor (here we consider infinite temperature $\beta=0$ first for simplicity and the finite temperature case will be analyzed in the end of this section) is given by
\begin{equation}
    g^{con}(s)=\braket{\text{tr} (\rho_A^{is})\text{tr}(\rho_A^{-is})}_c
\end{equation}
Depending on whether $\lambda\leq 1$ or $\lambda\geq 1$ we can respectively analytically continue Equation~\ref{correlator} or Equation~\ref{correlator2} to obtain $\tilde g(s)$, as they take the same form except for the overall factor, we will focus on the $\lambda\leq 1$ case from now on. After analytical continuation we have
\begin{equation}\label{ramp}
    g^{con}(s)=\sum_{c=1}^{\infty} f(c,s)
\end{equation}
where $f(c,s)$ is defined as
\begin{equation}\label{rampseries}
    f(c,s)=\lambda^c\binom{-is}{c}\binom{is}{c}
    {}_2F_1(\substack{-is,-is+c\\ c+1}|\lambda){}_2F_1(\substack{is,is+c\\ c+1}|\lambda)
\end{equation}
Here we have kept the $\lambda$ dependence implicit. This expression is of order $\frac{1}{N_A^2}$ compared to the leading term (see the discussion of the disconnected part below), which agrees with our expectation for the ramp. Next we will show that $f(c,s)$ decays exponentially for large $c$ and therefore the sum~\ref{ramp} converges.\\ \\
To prove that Equation~\ref{ramp} gives the linear ramp, we start from the observation that the hypergeometric functions appearing in the sum can be written in terms of the following integral representation
\begin{equation}\label{integralrepresentation}
    {}_2F_1(\substack{a,b\\c}|\lambda)=\frac{\Gamma(c)\Gamma(1+b-c)}{2\pi i \Gamma(b)}
    \int_0^{(1+)}\frac{t^{b-1}(t-1)^{c-b-1}}{(1-\lambda t)^a}dt
\end{equation}
which is valid as long as $\arg(1-\lambda)<\pi$, $c-b\neq 1,2,3\ldots$ and $\text{Re}(b)>0$. Here the notation $\int_0^{(1+)}$ denotes a specific contour which starts at $0$, circles $1$ once in the positive direction and returns to $0$, while not touching the branch cut $(\frac{1}{\lambda},\infty)$. Since we are sticking to the case where $c$ is an integer, the function in the integral does not have a branch cut along the negative real axis, thus we can deform our contour into arbitrary closed curves which encircles the branch cut $[0,1]$ without intersecting $(\frac{1}{\lambda},\infty)$. Plugging this into Equation~\ref{rampseries} and expand the binomial coefficients in terms of gamma functions we find that
\begin{equation}
    f(c,s)=\frac{c\lambda^c}{4\pi^2}|I(c,s)|^2
\end{equation}
where
\begin{equation}
    I(c,s)=\int_0^{(1+)} dt\;t^{c-1+is}(t-1)^{-is}(1-\lambda t)^{-is}
\end{equation}
Now we are interested in the asymptotic behaviour of $I(c,s)$ at large $s$, which can be analyzed using the steepest descent method. To do this we define the function
\begin{equation}\label{exponent}
    G(t)=\alpha \ln t+i\ln\frac{t}{(t-1)(1-\lambda t)}
\end{equation}
where $\alpha=\frac{c-1}{s}>0$, now we can write $I(c,s)$ as
\begin{equation}\label{integral}
    I(c,s)=\int_0^{(1+)}
    dt\; e^{sG(t)}
\end{equation}
which is the starting point of the steepest descent analysis. $G(t)$ has two saddles given by
\begin{equation}\label{saddle}
    t_{1,2}=\frac{\alpha(\lambda+1)\pm\sqrt{\alpha^2(\lambda-1)^2-4\lambda}}{2(\alpha-i)\lambda}
\end{equation}
Note that the term in the square root changes sign at $\alpha^*=\frac{2\sqrt \lambda}{1-\lambda}$. A detailed analysis of the global property of $G(t)$ (See Appendix~\ref{append:saddle}) implies that for $\alpha<\alpha^*$, we have to deform the integration contour to pick up both $t_1$ and $t_2$ and their contributions are of the same order as we have $\text{Re}[G(t_1)]=\text{Re}[G(t_2)]$ in this case. The steepest descent method then gives
\begin{equation}\label{intasymp1}
    I(c,s)\sim\lambda^{-\frac{1}{2}(c-1)} \bigg\{
    \sqrt{\frac{2\pi}{s|G''(t_1)|}}e^{i\phi_1+is\text{Im}[G(t_1)]}
    +\sqrt{\frac{2\pi}{s|G''(t_2)|}}e^{i\phi_2+is\text{Im}[G(t_2)]}
    \bigg\}
\end{equation}
where $\phi_{1,2}$ are the two angles determined by the directions of the steepest descent paths at $t_{1,2}$. Plugging this into the expression of $f(c,s)$ gives
\begin{equation}\label{asymp1}
    f(c,s)\sim\frac{c\lambda}{4\pi^2 s} \bigg\{
    \frac{2\pi}{|G''(t_1)|}+\frac{2\pi}{|G''(t_2)|}+\frac{4\pi}{\sqrt{|G''(t_1)G''(t_2)|}}
    \cos(s\Delta\theta+\Delta \phi)
    \bigg\}
\end{equation}
where $\Delta\theta=\text{Im}[G(t_1)-G(t_2)]$ and $\Delta\phi=\phi_1-\phi_2$, they are both functions only of $\alpha$. On the contrary, when $\alpha>\alpha^*$, it can be shown that we only need to deform our integration contour to pick up $t_2$, and $t_1$ no longer contributes to the integral, this in turn gives the following expression for $I(c,s)$
\begin{equation}\label{intasymp2}
    I(c,s)\sim\sqrt{\frac{2\pi}{s|G''(t_2)|}}e^{i\phi_2+s\text{Re}[G(t_2)]+is\text{Im}[G(t_2)]}
\end{equation}
or 
\begin{equation}\label{asymp2}
    f(c,s)=\frac{c\lambda^c}{4\pi^2} \frac{2\pi}{s|G''(t_2)|} e^{2s\text{Re}[G(t_2)]}
\end{equation}
Since $\text{Re}[G(t_2)]$ is bounded from above, as shown in Figure~\ref{Regt_2}, $f(c,s)$ decays as $\lambda^c$ for large enough $c$, and thus the sum over $c$ converges. In Figure~\ref{behaviour of f} we show $f(c,s)$ for a few different values of $\lambda$.
Combining the results above we see that for large $s$ the sum~\ref{ramp} is dominated by terms with $\alpha<\alpha^*$, or $c<\alpha^*s+1$, the number of such terms scales as $\alpha^*s$.  
According to Equation~\ref{asymp1} these terms are linear in $c$ and contains a factor $\frac{1}{s}$, thus we expect the sum to be linear in $s$. In the following we give a strict proof. \\ \\
When $s$ is large, to the leading order in $s$, we can approximate the sum with the following integral
\begin{equation}\label{integral ramp}
    g^{con}(s)=\sum_{c=1}^{\infty}f(c,s)\sim\lambda s\int_0^\infty d\alpha f(\alpha s,s) 
\end{equation}
Note that $G''(t_1)=G''(t_2)=0$ when $\alpha=\alpha^*$ and $t_1=t_2=t^*$, therefore the approximations~\ref{asymp1} and~\ref{asymp2} break down near this critical point. The breakdown can be estimated to take place when $|sG''(t_{1,2})|\sim 1$. Thus we divide the integral into three parts: 
\begin{equation}
    g^{con}(s)=R_1(s)+R_2(s)+R_3(s)
\end{equation}
where the three parts are respectively integrated over $[0,\alpha^*-\delta]$, $[\alpha^*-\delta,\alpha^*+\delta]$ and $[\alpha^*+\delta,\infty]$. Here $\delta$ roughly measures the width of the breakdown region. It turns out that (see Appendix~\ref{append:integral} for details) only the first term contributes at the leading order, which is linear in $s$. And the final expression for the ramp is given by 
\begin{equation}\label{linearramp}
    g^{con}(s)\sim K(\lambda)s+\text{subleading terms}
\end{equation}
where 
\begin{equation}
    K(\lambda)=\frac{\lambda}{2\pi}\int_0^{\alpha^*} d\alpha\; \alpha
    \bigg\{
    \frac{1}{|G''(t_1)|}+\frac{1}{|G''(t_2)|}
    \bigg\}
\end{equation}
Thus we have proved the existence of the linear ramp with a formula for its amplitude. Illustrations of numerical results are shown in Figure~\ref{numerics for ramps}. 

\begin{figure}
	\centering
		\includegraphics[width=0.8\textwidth]{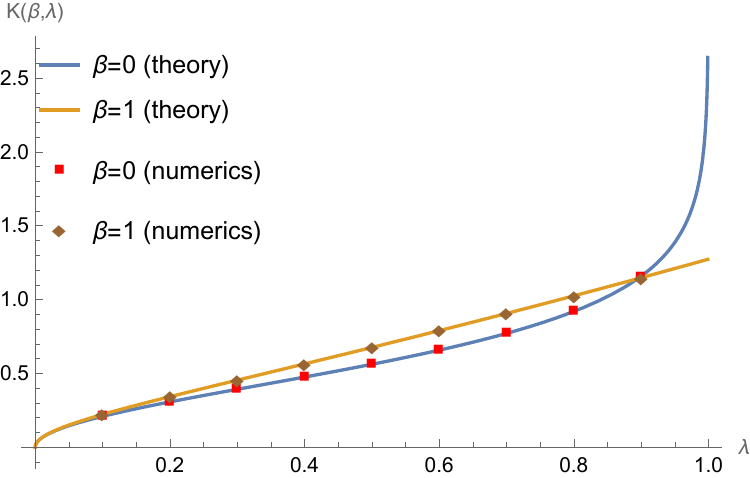}
        \captionsetup{format=hang}
	\caption{The amplitude of the ramp $K(\beta,\lambda)$, with both theoretical 
        and numerical results in nice agreement. Note that the $K(\beta,\lambda)$ diverges as $\lambda\rightarrow 1$ at $\beta=0$, and introducing $\beta>0$ cures this divergence, as expected from the calculation at $\lambda=1$} 
        \label{slope}
\end{figure}

\subsection*{Finite Temperature}
Generally we can consider the ramp at non-infinite temperature, this amounts to calculate 
\begin{equation}
    g^{con}_\beta(s)=\braket{\text{tr} (\rho_A^{\beta+is})\text{tr}(\rho_A^{\beta-is})}_c
\end{equation}
Here we will assume that $\beta\sim O(1)\ll s$. Now Equation~\ref{ramp} is modified to 
\begin{align}\label{finite T ramp}
    g^{con}_\beta(s)&=\frac{1}{N_A^{2\beta}}\sum_{c=1}^{\infty} f_\beta(c,s)
\end{align}
where
\begin{equation}
    f_\beta(c,s)=c\lambda^c\binom{-is+\beta}{c}\binom{is+\beta}{c}
    {}_2F_1(\substack{-is-\beta,-is-\beta+c\\ c+1}|\lambda){}_2F_1(\substack{is-\beta,is-\beta+c\\ c+1}|\lambda)
\end{equation}
Again we have suppressed the $\lambda$ dependence for simplicity. Note that $f_0(c,s)=f(c,s)$ defined above. In order to use the integral representation~\ref{integralrepresentation} we require $c-\beta>0$, this is not satisfied for small $c$'s. However under the assumption that $\beta$ is fixed and $\beta\sim O(1)$ we can ignore the contribution from $c<\beta$ terms as they vanish for large $s$~\cite{jones2001asymptotics}. In this case we have
\begin{align}\label{finite T ramp sum}
    g^{con}_\beta(s)=\frac{1}{N_A^{2\beta}}\sum_{c=[\beta]+1}^{\infty}\frac{c\lambda^c}{4\pi^2}|I_\beta(c,s)|^2
\end{align}
where
\begin{equation}
    I_\beta(c,s)=\int_0^{(1+)} dt\;F(\beta,t,\lambda)\;t^{c-1+is}(t-1)^{-is}(1-\lambda t)^{-is}
\end{equation}
in which
\begin{equation}
    F(\beta,t,\lambda)=\frac{(t-1)^\beta(1-\lambda t)^\beta}{t^\beta}
\end{equation}
For simplicity we will write $F(\beta,t,\lambda)$ as $F(t)$ in the following. The integrals in Equation~\ref{finite T ramp sum} can still be approximated using steepest descent method, which gives
\begin{equation}\label{finite temperature asymp1}
    I_\beta(c,s)\sim\lambda^{-\frac{1}{2}(c-1)} \bigg\{
        \sqrt{\frac{2\pi}{s|G''(t_1)|}}F(t_1)e^{i\phi_1+is\text{Im}[G(t_1)]}
        +\sqrt{\frac{2\pi}{s|G''(t_2)|}}F(t_2)e^{i\phi_2+is\text{Im}[G(t_2)]}
        \bigg\}
\end{equation}
for $\alpha<\alpha^*$, and 
\begin{equation}\label{finite temperature asymp2}
    I_\beta(c,s)\sim\sqrt{\frac{2\pi}{s|G''(t_2)|}}F(t_2)e^{i\phi_2+s\text{Re}[G(t_2)]+is\text{Im}[G(t_2)]}
\end{equation}
Following the same reasoning as in the $\beta=0$ case, we again find a ramp with the amplitude 
\begin{equation}
    K(\beta,\lambda)=\frac{\lambda}{2\pi N_A^{2\beta}}\int_0^{\alpha^*} d\alpha\; \alpha
    \bigg\{
    \frac{|F(t_1)|^2}{|G''(t_1)|}+\frac{|F(t_2)|^2}{|G''(t_2)|}
    \bigg\}
\end{equation}
And again for the $N_A>N_B$ case we redefine $\lambda\rightarrow\frac{1}{\lambda}$, and replace $N_A^{2\beta}\rightarrow N_B^{2\beta}$. As shown in Figure~\ref{slope}, introducing a non-zero $\beta$ cures the divergence of $K(\beta,\lambda)$ as $\lambda\rightarrow 1$. We will come back to this point again in the discussion of the critical case $\lambda=1$ below.

\subsection*{Disconnected Part}
To compare with the connected part discussed above we also need the disconnected part of the correlator, $g_d(\beta,s)$, which is given by
\begin{equation}
    g^{disc}_\beta(s)=|\braket{\text{tr} \rho_A^{\beta+is}}|^2
\end{equation}
We again do this by calculating $\braket{\text{tr} \rho_A^{n}}$ first and analytically continue $n$ to complex numbers. It is well known that the trace is given by summing over all disk non-crossing permutations at the leading order in $N_{A/B}$, the result is~\cite{kudler2021relative}
\begin{equation}
    \braket{\text{tr} \rho_A^{n}}=N_A^{1-n}{}_2F_1(\substack{1-n,-n\\ 2}|\lambda)
\end{equation}
here again $\lambda=\frac{N_A}{N_B}$, which is assumed to be smaller than $1$, if this is not the case we can use $\lambda'=\frac{N_B}{N_A}$ as before. Analytical continuation then gives
\begin{equation}
    g_\beta^{disc}(s)=N_A^{2-2\beta} |{}_2F_1(\substack{1-\beta-is,-\beta-is\\ 2}|\lambda)|^2
\end{equation}
With variable transformations and asymptotic formulas for ${}_2F_1$~\cite{lozier2003nist,jones2001asymptotics}, it is easy to show that 
\begin{equation}
    g^{disc}_\beta(s)\sim \frac{N_A^2}{N_A^{2\beta}}\frac{1}{s^3}
\end{equation}
which decays as $\frac{1}{s^3}$ and is larger than $g_c$ by order $N_A^2$, agreeing with the results in~\cite{chen2018universal,cotler2017black}.

\subsection*{Critical Case: $\lambda=1$}
It is worth considering the critical value $\lambda=1$ since this is the critical point where the subsystem $A$ is exactly half of the whole system. In this case the sum~\ref{sum1} can be done explicitly since it is just the total number of all connected annular non-crossing permutations in $\text{ANC}(n,m)$, which gives 
\begin{equation}
      \braket{\text{tr}\rho_A^n\;\text{tr}\rho_A^m}_c=\frac{1}{N_A^{m+n}}\frac{2nm}{n+m}\binom{2n-1}{n}\binom{2m-1}{m}
\end{equation}
Now we can readily do the analytical continuation by expanding the binomial coefficients in terms of gamma functions. In doing so it is useful to use the following Legendre duplication formula to simplify the expression
\begin{equation}
     \frac{\Gamma(2z)}{\Gamma(z)}=\frac{1}{2^{1-2z}\sqrt{\pi}}\Gamma(z+\frac{1}{2})
\end{equation}
After some algebra we get
\begin{equation}
    \braket{\text{tr}\rho_A^n\;\text{tr}\rho_A^m}_c=\frac{1}{N_A^{m+n}}\frac{2nm}{n+m}\cdot \frac{4^{m+n-1}}{\pi}\frac{\Gamma(n+\frac{1}{2})}{\Gamma(n+1)}
    \frac{\Gamma(m+\frac{1}{2})}{\Gamma(m+1)}
\end{equation}
The analytical continuation amounts to the substitution $n,m\rightarrow \beta\pm is$, and again we assume that $\beta\ll s$:
\begin{equation}\label{critical ramp}
    g^{con}_\beta(s)\sim \frac{4^{2\beta}}{4\pi N_A^{2\beta}}\cdot \frac{s}{\beta}
\end{equation}
where we have used $\Gamma(x+\alpha)\sim x^\alpha \Gamma(x)$ for large $s$. Note that a non-zero $\beta$ is necessary for the convergence, this agrees with our result above that including a non-zero $\beta$ cures the divergence of $K(\beta,\lambda)$ as $\lambda\rightarrow 1$\footnote{Another way to understand the divergence at $\lambda=1$ is to do the analytical continuation before summing up the series, using the gamma function representation of hypergeometric functions at $\lambda=1$, it is easy to see that for large but fixed $s$, $f(c,s)$ then decays as $\frac{1}{c}$ for large $c$, thus the sum becomes divergent.}.\\ \\
For the disconnected part, we again write the hypergeometric functions in terms of gamma functions, after applying the duplicate formula and taking the asymptotic limit, we have
\begin{equation}
    g^{disc}(\beta,s)=\frac{4^{2\beta}N_A^2}{N_A^{2\beta}} \frac{1}{\pi s^3}
\end{equation}
The results above, especially the scaling form $\frac{s}{\beta}$ of Equation~\ref{critical ramp}, takes the same form as the result in~\cite{saad2019jt}, where the ramp for a random matrix Hamiltonian is obtained by calculating in the dual JT gravity. The extra factor $N_A^{-2\beta}$ merely comes from our normalization $\text{tr}\rho_A=1$ and remains the same at all orders. So we can simply ignore it as an overall normalization constant. In the next section we will include more discussions along this line, where we will find a structure out of our annular non-crossing permutations similar to that appeared in the JT gravity calculation in~\cite{saad2019jt}.\\ \\
It is interesting to make a connection between our calculation using non-crossing permutations and the standard random matrix calculation. The spectral form factor is completely fixed by the connected correlator $\braket{\rho(E)\rho(E')}_{con}$ where $\rho(E)$ is the density of states at energy $E$. It can be shown that this correlator generally takes the form of a sine-kernel, which only depends on the symmetry type of the random matrix ensemble~\cite{brezin1993universality,nagao1992correlation}. In our case we are calculating the spectral form factor of $H=-\ln(X^\dagger X)$ where $X^\dagger X$ is a Wishart random matrix. It is easy to see that the ensemble of $H$ is invariant under unitary transformations and there is no other discrete symmetry, so it is still in the Gaussian unitary ensemble (GUE). For this symmetry type the sine kernel takes the form 
\begin{equation}
    \braket{\rho(E)\rho(E')}_{con}=\braket{\rho(\bar E)}\delta(\omega)-\braket{\rho(E)}\braket{\rho(E')}\frac{\sin^2[\pi\rho(\bar E)\omega]^2}{[\pi\rho(\bar E)\omega]^2}
\end{equation}
where we have defined $\bar E=\frac{(E+E')}{2}$ and $\omega=E-E'$. By repeating the calculation in~\cite{chen2018universal} for our ensemble of $H$ we find that the ramp is fixed by the edges of the spectrum of $H$, which is determined purely by the ratio $\lambda$. At infinite temperature the slope of the ramp is given by
\begin{equation}
    K(\lambda)=\frac{1}{\pi}\ln\frac{1+\sqrt{\lambda}}{1-\sqrt{\lambda}}
\end{equation}
while for finite temperature we have 
\begin{equation}
    K(\beta,\lambda)=\frac{(1+\sqrt{\lambda})^{4\beta}-(1-\sqrt{\lambda})^{4\beta}}{4\pi\beta}
\end{equation}
These expressions agree numerically with the results we obtained using combinatorics, reaffirming the equivalence between the two approaches. It is also interesting to discuss the relationship between the slope of the ramp and the symmetry type of the random matrix ensemble. From the discussion above we see that the sine kernel is determined by the symmetry type as well as $\rho(E)$ which is the density of states. The latter is non-universal, therefore the slope is not uniquely determined by the symmetry type. However, for a certain random matrix ensemble, we can rescale the energy to make $\rho(E)$ constant, which is known as unfolding~\cite{haake1991quantum}. In this case the ramp can be uniquely fixed by the symmetry type.

\section{Number of Connected Loops as Geodesic Length?}
\label{sec:geodesic}
\begin{figure}
	\centering
	\begin{subfigure}{0.5\textwidth}
		\includegraphics[width=\textwidth]{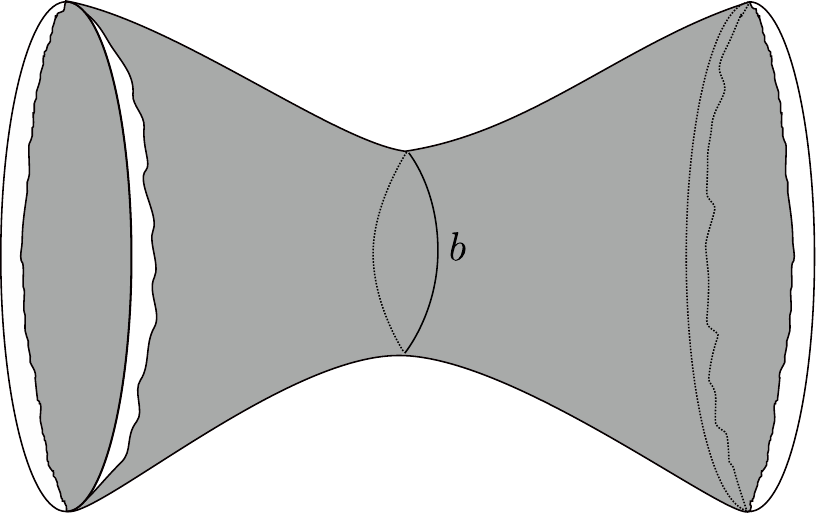}
        \captionsetup{format=hang}
		\caption{The double trumpet geometry}
        \label{double cone}
        \vspace*{0.7cm}
	\end{subfigure}
	\begin{subfigure}{.7\textwidth}
		\includegraphics[width=\textwidth]{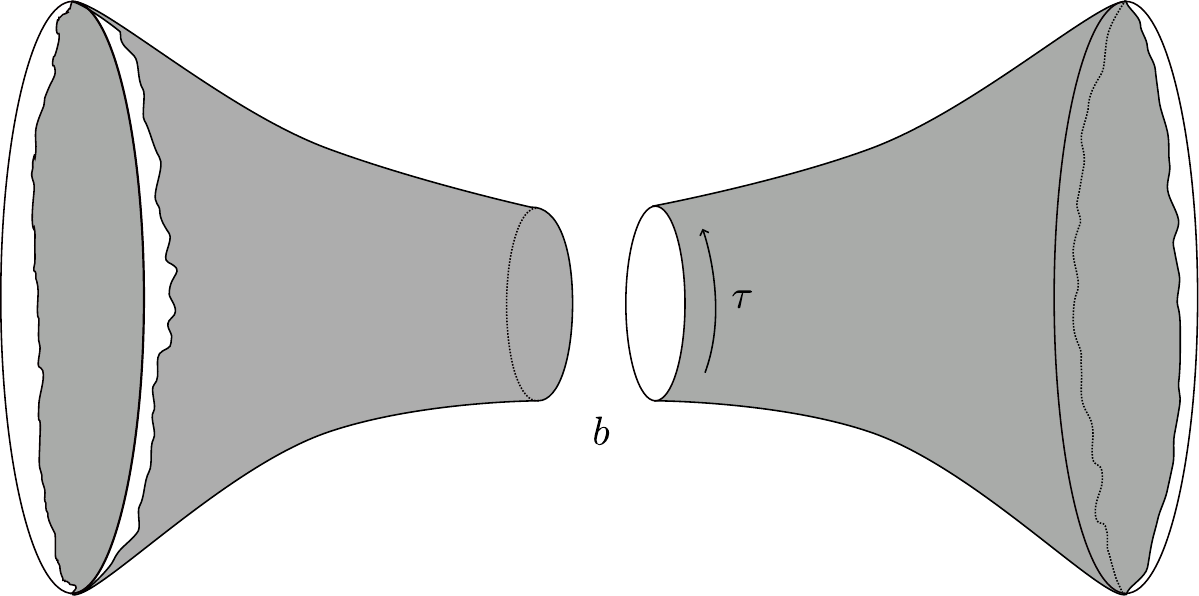}
        \captionsetup{format=hang}
        \caption{Gluing two single trumpets together}
	\label{gluing two trumpets}
        \end{subfigure}
        \captionsetup{format=hang}
        \caption{The bulk geometry in the calculation of $\braket{Z(\beta_1)Z(\beta_2)}_c$}
        \label{double trumpet}
\end{figure}

\begin{figure}
	\centering
	\begin{subfigure}{0.4\textwidth}
		\includegraphics[width=\textwidth]{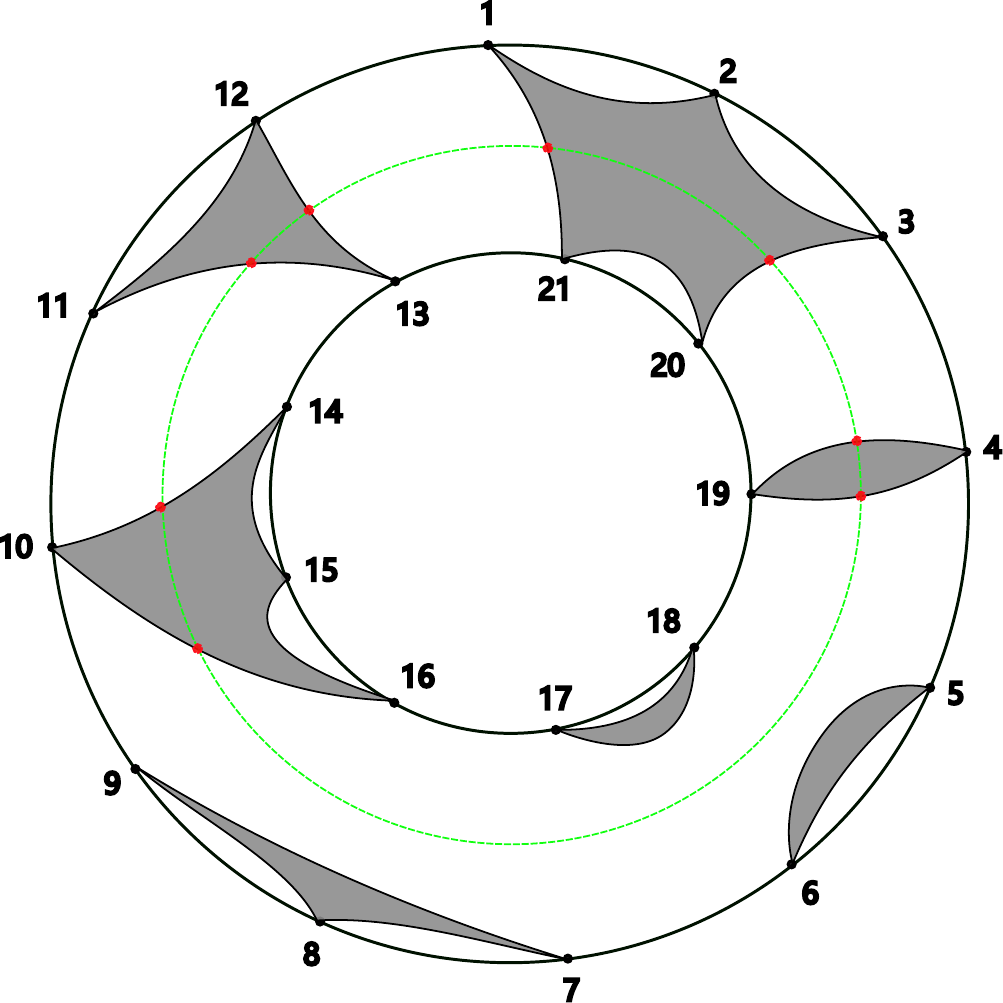}
        \captionsetup{format=hang}
		\caption{The cut intersects each connected orbit at two adjacent points (red)}
        \label{cut ANC}
        \vspace*{0.7cm}
	\end{subfigure}
        \hspace{3em}
	\begin{subfigure}{0.4\textwidth}
		\includegraphics[width=\textwidth]{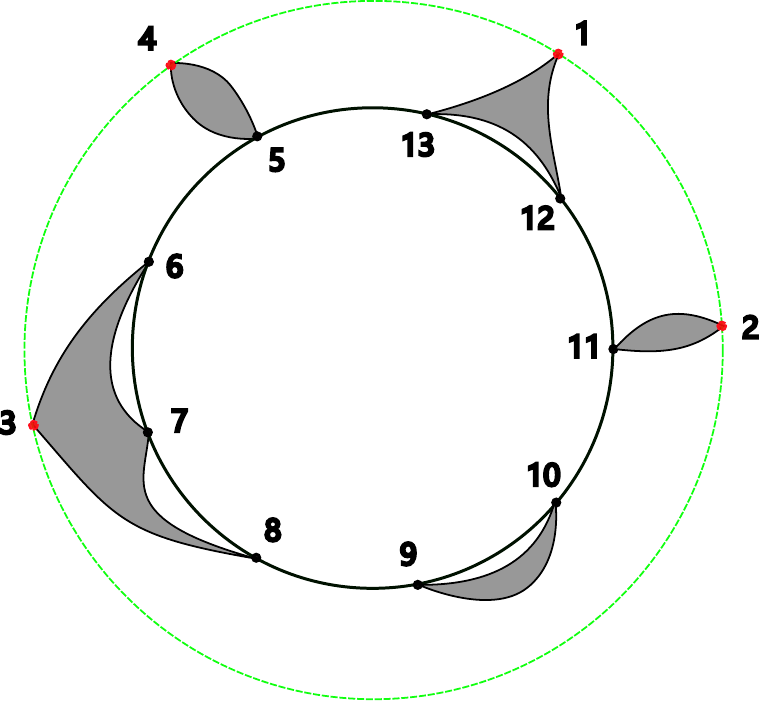}
        \captionsetup{format=hang}
        \caption{Permutation $\tau_1$ induced by the cut}
	\label{inner ANC}
        \vspace*{1.2cm}
        \end{subfigure}
        \begin{subfigure}{0.4\textwidth}
		\includegraphics[width=\textwidth]{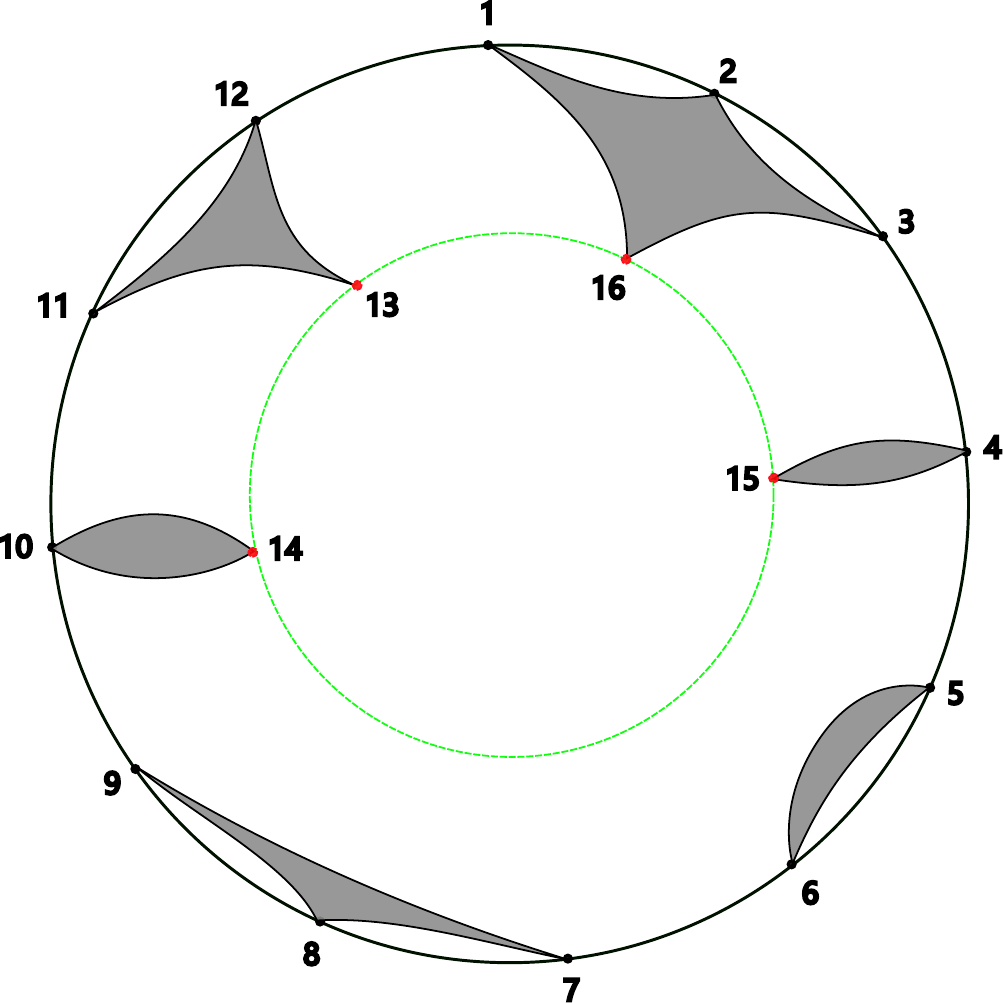}
        \captionsetup{format=hang}
        \caption{Permutation $\tau_2$ induced by the cut}
	\label{outer ANC}
        \end{subfigure}
        \captionsetup{format=hang}
        \caption{Slicing open an annular non-crossing permutation $\tau$ induces two new annular-non-crossing permutations, denoted $\tau_1$ and $\tau_2$. When we try to glue $\tau_1$ and $\tau_2$ together, there is an ambiguity associated to a relative shift between bulk elements of $\tau_1$ and $\tau_2$. For example, we can glue them via the identification $1_{\tau_1}\leftrightarrow 16_{\tau_2},2_{\tau_1}\leftrightarrow 15_{\tau_2},3_{\tau_1}\leftrightarrow 14_{\tau_2},4_{\tau_1}\leftrightarrow 13_{\tau_2}$, which restores $\tau$. However, we can also introduce a relative shift between them before gluing, such as $2_{\tau_1}\leftrightarrow 16_{\tau_2},3_{\tau_1}\leftrightarrow 15_{\tau_2},4_{\tau_1}\leftrightarrow 14_{\tau_2},1_{\tau_1}\leftrightarrow 13_{\tau_2}$, which reproduces some $\tau'\neq\tau$.}
        \label{slicing ANC}
\end{figure}

Above we have calculated the correlator $\braket{\text{tr}\rho_A^n\;\text{tr}\rho_A^m}_c$ as well as its analytical continuation by making connections to annular non-crossing permutations. In this section we view these permutations from another perspective, which bears some similarity to the results in~\cite{saad2019jt}, possibly indicating relationships between the two. \\ \\
In~\cite{saad2019jt}, the bulk dual of the correlator $\braket{Z(\beta_1)Z(\beta_2)}_c$ was calculated, where $Z(\beta)=\text{tr} e^{-\beta H}$ with $H$ a random matrix from certain ensembles. It turns out that the corresponding bulk geometry, to the first order in $\frac{1}{G}$, is a double trumpet, as shown in Figure~\ref{double cone}.
Despite two wiggling boundaries with lengths set by $\beta_{1,2}$, this double trumpet geometry has two independent parameters, the first is the geodesic length $b$ at the throat of the geometry. The other parameter can be understood by thinking of the double trumpet as two single trumpets glued together along the geodesic at the throat, as shown in Figure~\ref{gluing two trumpets}. Thus we need to introduce a relative twist $\tau\in[0,b]$ between the two single trumpets before gluing. This leads to the following way to evaluate the bulk path integral: we first average over boundary wiggles for a single trumpet with fixed $b$ and $\beta$, which gives us a function $Z^{trumpet}_{Sch}(\beta,b)$ (here the lower index 'Sch' means that the boundary action is a Schwarzian). Then we take two copies of the single trumpet and do the gluing, in doing so we need to integrate over the twist $\tau$, which simply gives a factor $b$. Finally we integrate over $b$ to complete the path integral. Following the steps above we get
\begin{equation}\label{SSS result}
    \braket{Z(\beta_1)Z(\beta_2)}_c\sim \int bdb\; Z^{trumpet}_{Sch}(\beta_1,b) Z^{trumpet}_{Sch}(\beta_2,b)
\end{equation}
Interestingly, a very similar structure appears in the evaluation of $\braket{\text{tr}\rho_A^n\;\text{tr}\rho_A^m}_c$ by summing over annular non-crossing permutations. To see this, we slice up an annular non-crossing permutation $\tau$ with $c$ connected orbits (which we denote as $\tau\in\text{ANC}(n,m,c)$, as defined in appendix~\ref{append:annular}) into two halves by introducing a cut between the interior and exterior circles, as shown in Figure~\ref{slicing ANC}.
The definition of annular non-crossing permutations ensures that the cut intersects each connected loop at exactly a pair of adjacent points, as shown in Figure~\ref{cut ANC}. Thus we can merge these pairs into $c$ points on the cut. This procedure defines two new annular non-crossing permutations, we denote them respectively by $\tau_1\in\text{ANC}(c,m,c)$ and $\tau_2\in\text{ANC}(n,c,c)$, see Figure~\ref{inner ANC} and~\ref{outer ANC}. In the following we call elements on the cut \emph{bulk elements} and elements on the interior and exterior circles \emph{boundary elements}, as denoted by red and black dots in Figure~\ref{slicing ANC}. There is, however, a subtlety here: the cutting procedure does not induce a unique labeling for the bulk elements. Rather, the labeling is only defined up to a cyclic permutation. For example, for $\tau_1$ in Figure~\ref{inner ANC}, it is equally good if we relabel $1\rightarrow 2\rightarrow 3\rightarrow 4\rightarrow 1$, which results in a new annular non-crossing permutation $\tau_1'\neq \tau_1$. This ambiguity can be formulated in the following way: we start with the set $\text{ANC}(c,m,c)$ and the cyclic group $G_e$ generated by the permutation $\gamma_e=(1,2,3\ldots c)$, which shifts the $c$ bulk elements in a cyclic way. It is easy to see that $G_e$ is of order $c$ (that is, it contains $c$ different elements). Cyclic relabeling of bulk elements discussed above can then be realized by acting $G_e$ on $\text{ANC}(c,m,c)$ adjointly. 
\begin{equation}
    g\cdot \tau=g\circ\tau \circ g^{-1} 
\end{equation}
where $g\in G_e$ and $\circ$ is the composition of permutations. It is easy to verify that $g\cdot \tau\in\text{ANC}(c,m,c)$. Furthermore, this group action is free in the sense that $g\cdot \tau\neq \tau$ for $g\neq e$, where $e$ is the unit element of $G_e$. Now we can define \emph{orbits} of the set $\text{ANC}(c,m,c)$ under the group action of $G_e$ (not to be confused with orbits of a permutation): two elements $\tau_1$ and $\tau_1'$ are said to be in the same orbit if and only if there is some $g\in G_e$ such that $g\cdot \tau_1=\tau_1'$. This is an equivalence relation, the set of equivalence classes and its elements will be denoted as $\overline{\text{ANC}}(c,m,c)$ and $[\tau_1]$. The group action being free implies that the number of elements in $\overline{\text{ANC}}(c,m,c)$ is given by $\#\overline{\text{ANC}}(c,m,c)=\frac{\#\text{ANC}(c,m,c)}{c}$. We can similarly define the set of equivalence classes $\overline{\text{ANC}}(n,c,c)$ and its elements $[\tau_2]$, starting from the set $\text{ANC}(n,c,c)$ and the group $G_i$ generated by $\gamma_i=(n+1,n+2,\ldots, n+c)$. With these notations, the cutting procedure defined above induces two equivalence classes $[\tau_1]$ and $[\tau_2]$. On the other hand, we can reverse the cutting procedure to form an arbitrary $\tau\in\text{ANC}(n,m,c)$ by first choosing $[\tau_1]\in\overline{\text{ANC}}(c,m,c)$ and $[\tau_2]\in\overline{\text{ANC}}(n,c,c)$, then gluing the bulk elements of $[\tau_1]$ and $[\tau_2]$ in a cyclic way. However, this does not uniquely fix $\tau$. The subtlety is that we only require the bulk elements of $[\tau_1]$ and $[\tau_2]$ to be glued in a cyclic way, but we have not yet fixed the \emph{relative shift} between them, see Figure~\ref{cut ANC} for example. It is easy to see that there are $c$ possible values for the relative shift and each gives a different $\tau$.\\ \\
Now we are ready to view the enumeration of annular non-crossing permutations and evaluate Equation~\ref{correlator} from a new perspective. Again we divide the sum into sectors with different $c$. For fixed $c$, we first enumerate $[\tau_1]$. The enumeration of $[\tau_1]$ with a fixed number of interior orbits $s=0,1,\ldots,m$ can be obtained by simply dividing the respective enumeration of $\tau_1$ by a factor $c$, which is given by $\binom{m}{s}\binom{m}{s+c}$. Therefore, summing over $s$ gives a factor
\begin{equation}
    Z(m,c)=\binom{m}{c}{}_2F_1(\substack{-m,-m+c\\ c+1}|\lambda)
\end{equation}
Enumeration of $[\tau_2]$ and summation over all possible numbers $r$ of exterior orbits in $[\tau_2]$ gives another factor $Z(n,c)$. Then we glue them together and sum over all possible values of the relative shift, which gives a factor $c$. In this way we can recast Equation~\ref{sum1} as  
\begin{equation}
    \braket{\text{tr}\rho_A^n\;\text{tr}\rho_A^m}_c
    =\frac{1}{N_A^{n+m}}\sum_c c\lambda^c Z(m,c)Z(n,c)
\end{equation}
This sum takes a similar form as Equation~\ref{SSS result} except that we have an extra weight $\lambda^c$ here, which vanishes when $\lambda=1$. Interestingly it is also at this value of $\lambda$ that our result for the ramp has the same scaling behaviour as in~\cite{saad2019jt}. It should be noticed that our calculation bears a series of similarities to the JT gravity case: the number of connected loops plays the role of the geodesic length, while the relative shift between $\tau_1$ and $\tau_2$ plays the role of the twist. Summing over exterior and interior orbits can be seen as an analog of the average over wiggling boundaries. Based on these similarities, we propose that the annular non-crossing permutations may be viewed as a discretized version for some bulk gravity theory.

\section{Conclusion and Discussion}
\label{sec:conclusion}
In this paper we have implied the replica trick to the evaluation of the double trace correlator $\braket{\text{tr}\rho_A^n\;\text{tr}\rho_A^m}$ of a density matrix $\rho_A$ obtained by partially tracing a Haar random state. We explored the relationship between the double trace correlator and annular non-crossing permutations with physical interpretations from different perspectives, and we argued that in the AdS/CFT setup these permutations naturally arise from a set of non-trivial spacetime saddles in the bulk dual. We analytically continued the correlator to general complex values of $n,m$ to evaluate the spectral form factor of the corresponding modular Hamiltonian $K=-\ln \rho_A$ and identified a ramp, which serves as an indicator of chaotic behaviours. A universal formula for the amplitude of the ramp was derived by applying the steepest descent method to the spectral form factor. Our results takes a similar form to that obtained from random matrix theory and dual JT gravity calculations in earlier literature, possible relationships between the two had been discussed. \\ \\
There are a series of open questions which are left untouched in this paper. Firstly, we only discussed the leading order contribution ($\frac{1}{N^2}$ term) to the ramp, which gives the linear regime but not the plateau. In order to understand the full dip-ramp-plateau picture we need to sum up certain higher order contributions, which correspond to annular permutations with complicated topology~\cite{Saad:2022kfe}. These permutations cannot be drawn in a \emph{planar} (that is, non-crossing) way on an annul unless we add \emph{handles} to it. This is the natural analog of summing over bulk saddles with higher genus in the random matrix/JT gravity theory. Due to these similarities the relationship between the two is worth further exploration. Secondly, in~\cite{saad2018semiclassical} a Lorentzian geometry was obtained by complexifying the double trumpet geometry and a zero mode was found responsible for the ramp. In our case, although we have identified a series of similarities between annular non-crossing permutations and the double trumpet geometry, there is no natural way to define a corresponding Lorentzian geometry and the zero mode which gives the ramp. Another interesting question is to consider random matrix ensembles in other symmetry classes~\cite{zirnbauer2010symmetry}. It had been proposed in~\cite{stanford2019jt} that random matrix models with different symmetries are dual to different bulk geometries. For example, if the boundary random matrix ensemble has time reversal symmetry then the sum over bulk geometries will include unorientable manifolds. Therefore it is natural to ask what if we do our calculation using matrix ensembles with non-trivial symmetries and whether there is a non-crossing permutation analog of unorientable bulk manifolds. Finally, based on the formalism in this paper, it is natural to ask whether we can find a way to calculate the ramp for an arbitrary horizon in the holographic setup. We will leave these discussions for future works.

\subsection*{Acknowledgments}

We thank Mike Stone, Arvin Shahbazi-Moghaddam, Simon Lin and Marc Klinger for useful discussions. This research is supported in part by 
the Air Force Office of Scientific Research under award number FA9550-19-1-036 and by the DOE award number DE-SC0015655.

\appendix
\section{Annular Non-Crossing Permutations}
\label{append:annular}

\begin{figure}
	\centering
	\begin{subfigure}{.45\textwidth}
		\includegraphics[width=\textwidth]{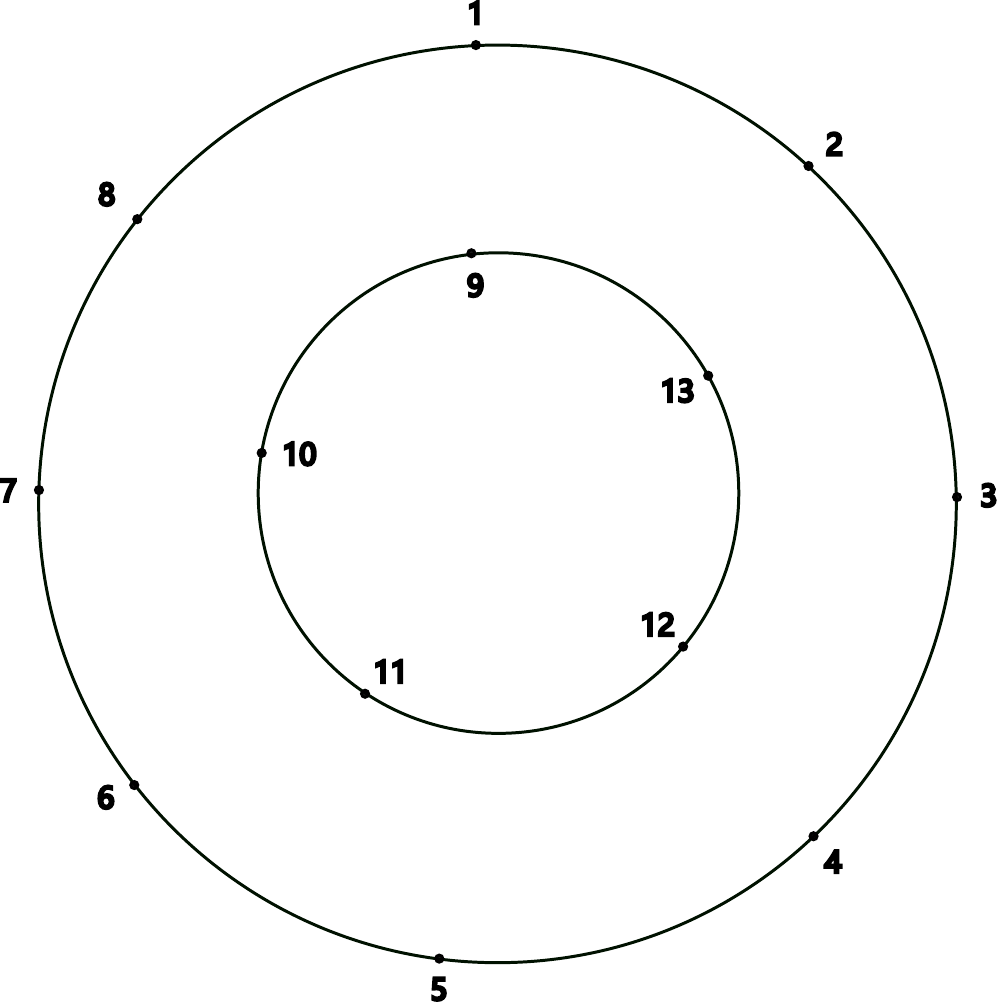}
        \captionsetup{format=hang}
		\caption{The exterior and interior circles with the example $n=8,m=5$}
        \label{two circles}
        \vspace*{0.7cm}
	\end{subfigure}
        \hspace{2em}
	\begin{subfigure}{.45\textwidth}
		\includegraphics[width=\textwidth]{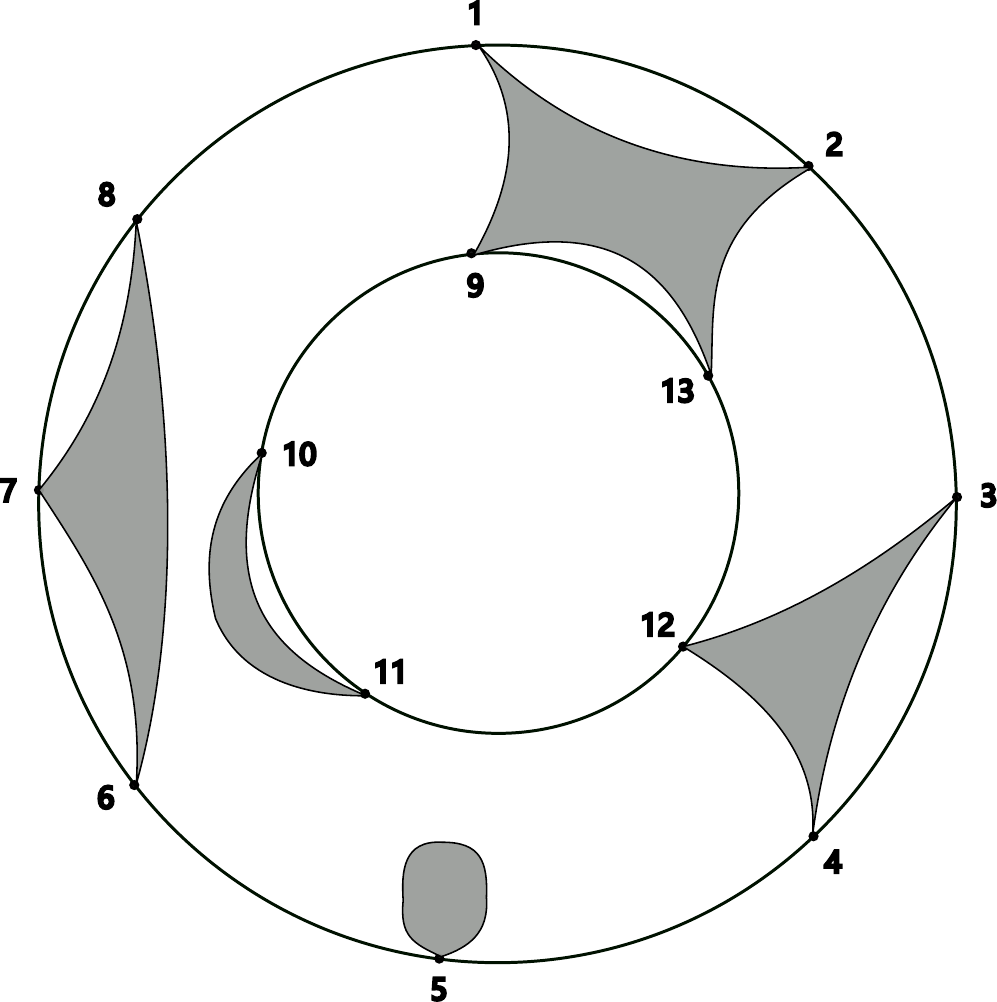}
        \captionsetup{format=hang}
        \caption{The annular non-crossing permutation $(1,2,13,9)(3,4,12)(5)(6,7,8)(10,11)$}
	\vspace*{0.7cm}
        \end{subfigure}
        \begin{subfigure}{.45\textwidth}
		\includegraphics[width=\textwidth]{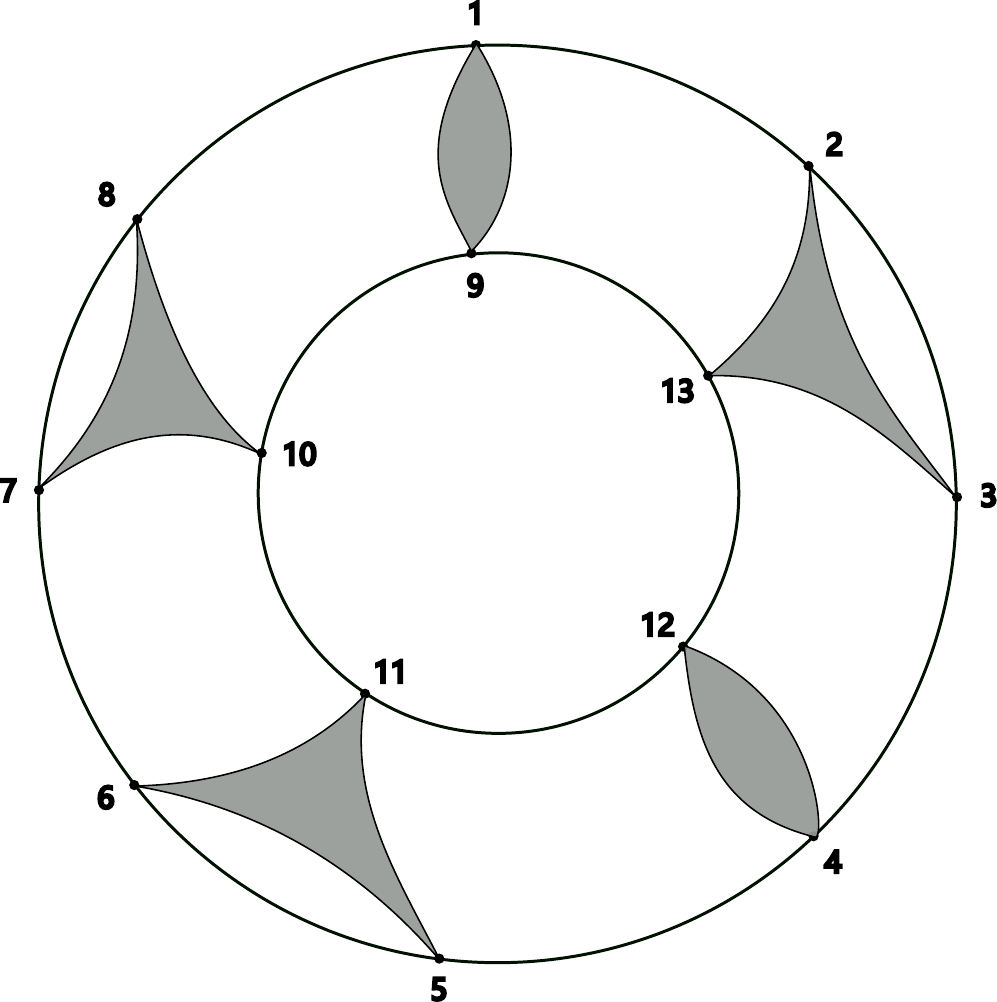}
        \captionsetup{format=hang}
        \caption{The annular non-crossing permutation $(1,9)(2,3,13)(4,12)(5,6,11)(7,8,10)$}
	\end{subfigure}
        \hspace{2em}
        \begin{subfigure}{.45\textwidth}
		\includegraphics[width=\textwidth]{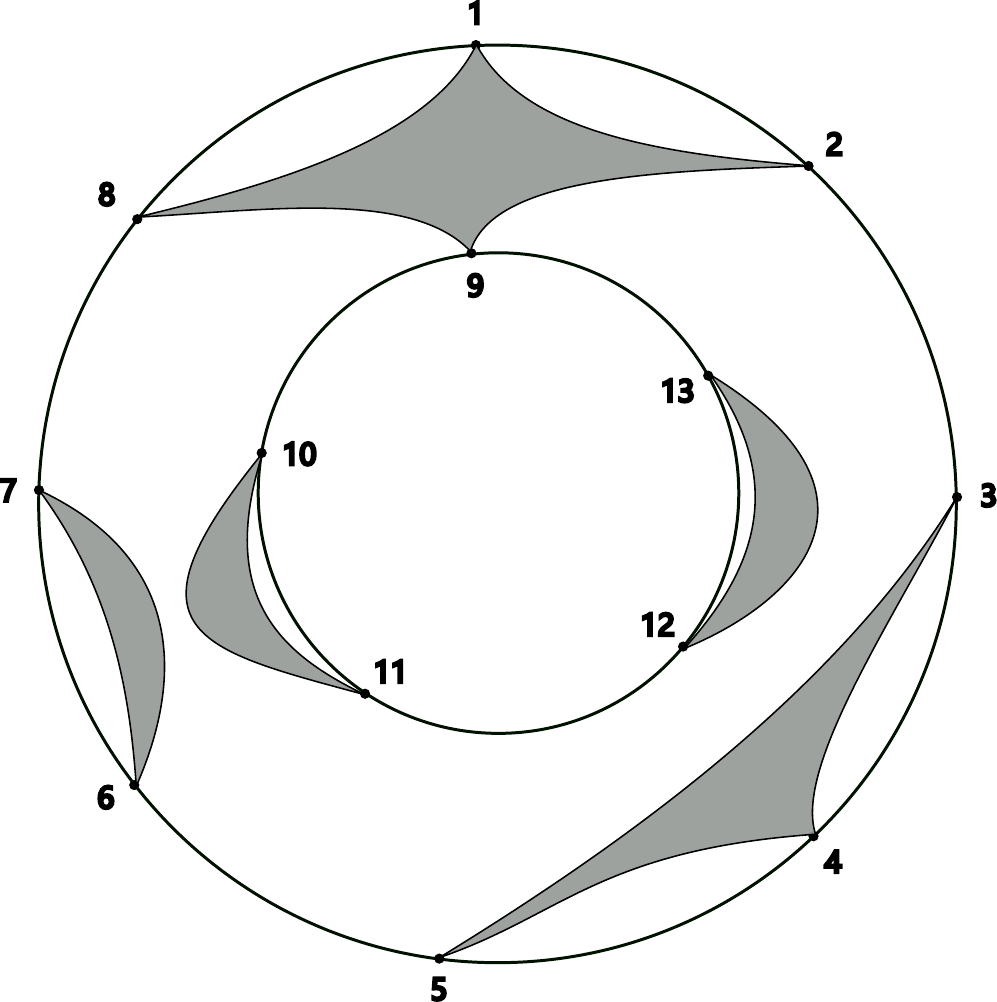}
        \captionsetup{format=hang}
        \caption{The annular non-crossing permutation $(1,2,8,9)(3,4,5)(6,7)(10,11)(12,13)$}
	\end{subfigure}
        \caption{Illustrations of annular non-crossing permutations}
        \label{ANC figure}
\end{figure}

In this appendix we briefly introduce some properties of annular non-crossing permutations which are used in the main text. Consider the permutation group $S_{n+m}$ where $n,m$ are two positive integers. We can illustrate this by drawing the first $n$ elements clockwise on an exterior circle and the remaining $m$ elements counterclockwise in an interior circle, as shown in Figure~\ref{two circles}. Each permutation $\tau\in S_{n+m}$ contains a number of orbits. An orbit is called \emph{exterior(interior)} if it only contains elements on the exterior(interior) circle, and \emph{connected} if it contains elements on both circles. The \emph{size} of an orbit is defined to be the number of elements in it. The \emph{exterior(interior) size} of a connected orbit is defined to be the number of exterior(interior) elements in it. Now let $e_k$ and $i_k$ respectively denote elements on the exterior and interior circle. If an orbit with exterior size $u$ and interior size $v$ can be written as $(e_re_{r+1}\ldots e_1 e_2\ldots e_{r-1} i_s i_{s+1}\ldots i_v i_1 i_2\ldots i_{s-1})$ where $r$ and $s$ are two integers which satisfy $1\leq r\leq u$, $1\leq s\leq v$, then the orbit is called \emph{clockwise oriented}. Pictorially, for any permutation $(b_1b_2b_3\ldots)$, we can connect $b_k$ to $b_{k+1}$ with a (directed) line and thus each clockwise oriented orbit can be represented by a (clockwise directed) polygon, see Figure~\ref{ANC figure}. If a permutation $\tau$ satisfies 
\begin{enumerate}
    \item All its orbits are clockwise oriented.
    \item There is no intersection between polygons for different orbits
\end{enumerate}
Then $\tau$ is said to be \emph{annular non-crossing}. The set of all annular non-crossing permutations is denoted as $\text{ANC}(n,m)$. Examples of annular non-crossing permutations are shown in Figure~\ref{ANC figure}. 
Furthermore, an annular non-crossing permutation is called \emph{connected} if it contains at least one connected orbit.\\ \\
Here we list some results for the enumeration of annular non-crossing permutations~\cite{kim2013cyclic}. Since disconnected annular non-crossing permutations are just two copies of disk non-crossing permutations, we will focus on connected ones from now on. Before going on we need some definitions. We define $\text{Par}(n,k)$ as the set of integer partitions of $n$ with $k$ parts. An element $\mu\in \text{Par}(n,k)$ can be written as $(1^{m_1},2^{m_2},\ldots )$. Where $m_i$'s are the numbers of subsets with size $i$ (that is, with $i$ elements), with the constraint $\sum m_i=k$ and $\sum im_i=n$. Also, for $\mu\in\text{Par}(n,k)$ we define the multinomial coefficient to be
\begin{equation}
    \binom{k}{\mu}=\binom{k}{m_1,m_2,\ldots}\equiv\frac{k!}{m_1!m_2!\ldots}
\end{equation}
An annular non-crossing permutation is said to be of exterior orbit type $\mu\in\text{Par}(R,r)$ if it has $r$ exterior orbits with total size $R$, and the numbers of orbits with different sizes are given by $\mu$. It is said to be of exterior connected orbit type $\lambda\in\text{Par}(n-R,c)$ if it has $c$ connected orbits with a total exterior size $n-R$, and the numbers of connected orbits with different exterior sizes are given by $\lambda$. Similarly we can define interior orbit types and connected interior orbit types. Now we can define the following enumerations
\begin{itemize}
    \item $\#\text{ANC}(n,m)$ is the number of connected annular non-crossing permutations
    \item $\#\text{ANC}(n,m;c)$ is the number of $\tau\in\text{ANC}(n,m)$ with $c$ connected orbits
    \item $\#\text{ANC}(n,m;c,r,s)$ is the number of $\tau\in\text{ANC}(n,m)$ with $c$ connected orbits, $r$ exterior orbits and $s$ interior orbits. 
    \item $\#\text{ANC}(n,m;c,r,s,R,S)$ is the number of $\tau\in\text{ANC}(n,m)$ with $c$ connected orbits, $r$ exterior orbits and $s$ interior orbits, and the total size of exterior/interior orbits is R/S.
    \item $\#\text{ANC}(n,m;c,r,s,R,S,\alpha,\beta,\lambda,\mu)$ is the number of $\tau\in\text{ANC}(n,m)$ with exterior orbit type $\alpha\in\text{Par}(R,r)$, interior orbit type $\beta\in\text{Par}(S,s)$, exterior connected orbit type $\lambda\in\text{Par}(n-R,c)$ and interior connected orbit type $\mu\in\text{Par}(m-S,c)$. 
\end{itemize}
And they are given by the following formulas
\begin{align}
    \#\text{ANC}(n,m;c,r,s,R,S,\alpha,\beta,\lambda,\mu)&=\frac{(n-R)(m-S)}{c}\binom{n}{r}\binom{m}{s}\binom{r}{\alpha}\binom{s}{\beta}\binom{c}{\lambda}\binom{c}{\mu}\\
    \#\text{ANC}(n,m;c,r,s,R,S)&=c\binom{n}{r}\binom{m}{s}\binom{R-1}{r-1}\binom{S-1}{s-1}\binom{n-R}{c}\binom{m-S}{c}\\
    \#\text{ANC}(n,m;c,r,s)&=c\binom{n}{r}\binom{m}{s}\binom{n}{r+c}\binom{m}{s+c}\\
    \#\text{ANC}(n,m;c)&=c\binom{2n}{n-c}\binom{2m}{m-c}\\
    \#\text{ANC}(n,m)&=\frac{2mn}{m+n}\binom{2n-1}{n}\binom{2m-1}{m}
\end{align}
The last of these equations had been derived in~\cite{mingo2004annular} as well.

\section{Details of the Saddle Point Analysis}
\label{append:saddle}
\begin{figure}
	\centering
		\includegraphics[width=0.8\textwidth]{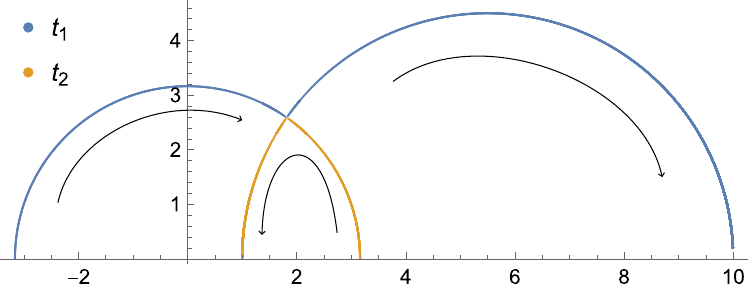}
        \captionsetup{format=hang}
        \caption{Saddle points for $\lambda=0.1$ and different values of $\alpha$ on the 
         complex plane, the blue and yellow contours are for $t_1$ and $t_2$ respectively, as we crank up $\alpha$ they move as shown by the arrows} 
        \label{figure for saddles}
\end{figure}

\begin{figure}
	\centering
	\begin{subfigure}{.55\textwidth}
		\includegraphics[width=\textwidth]{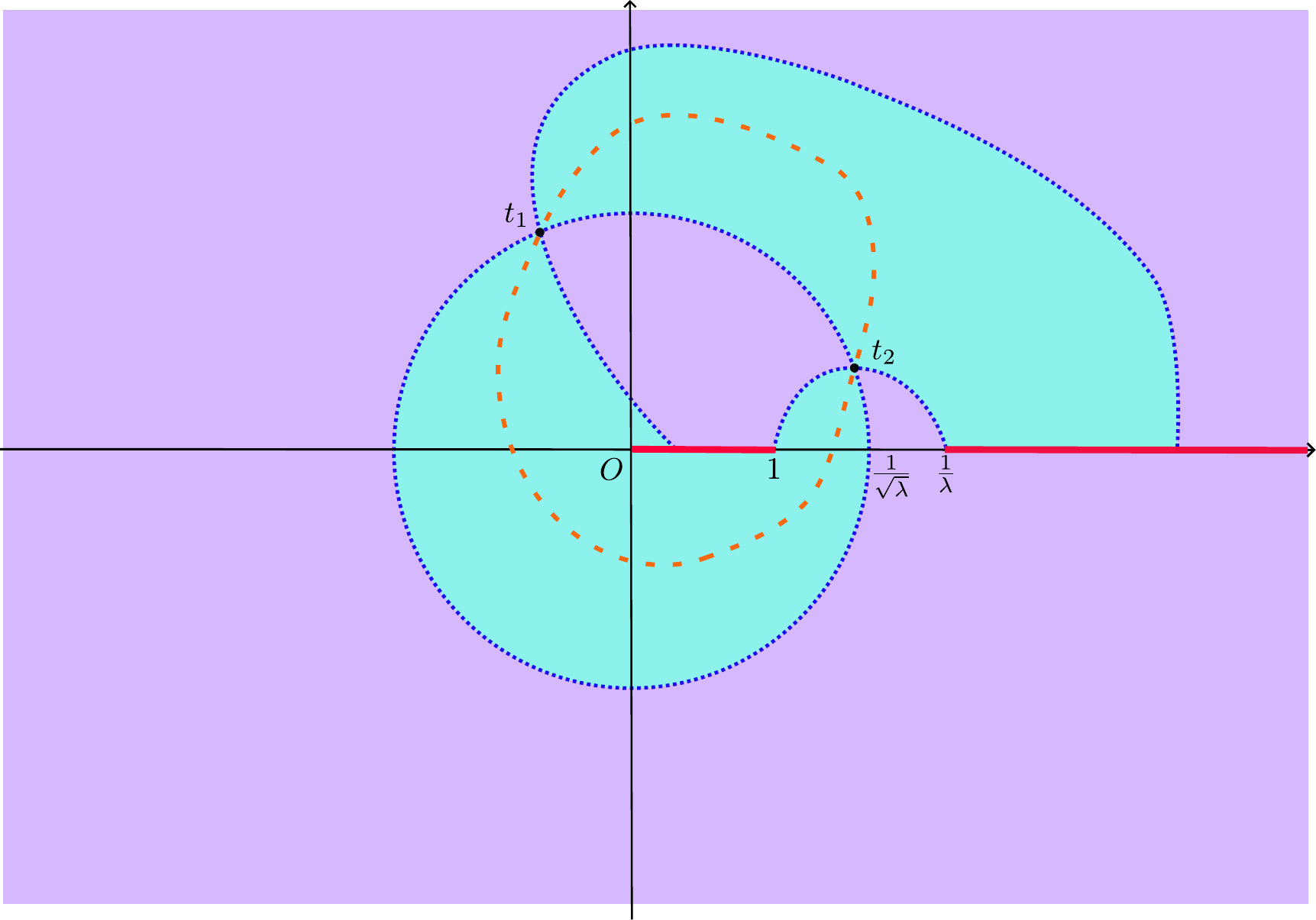}
        \captionsetup{format=hang}
		\caption{The integration contour for $\alpha<\alpha^*$}
        \label{cont1}
	\end{subfigure}
	\begin{subfigure}{.55\textwidth}
        \captionsetup{format=hang}
        \includegraphics[width=\textwidth]{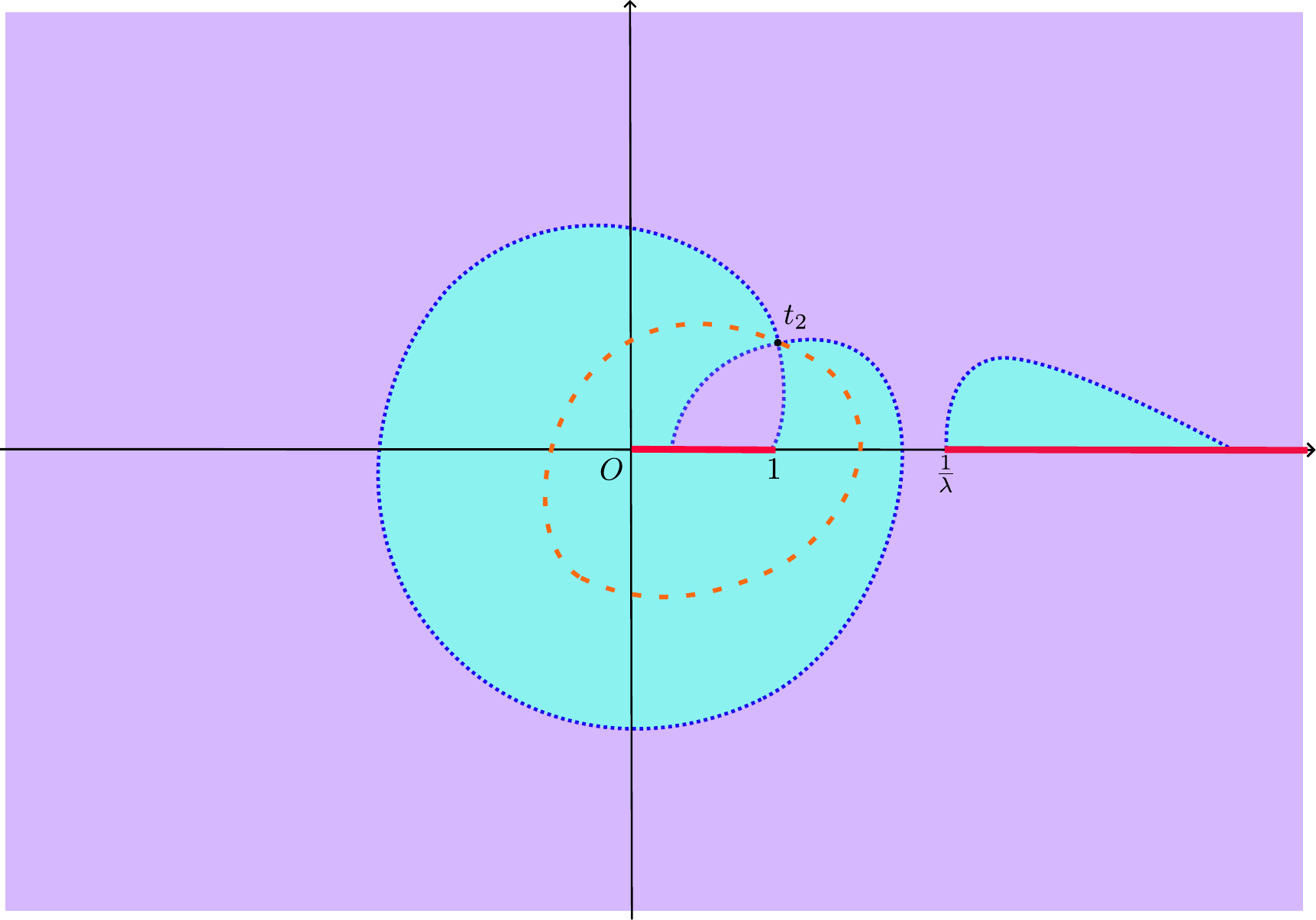}
		\caption{The integration contour for $\alpha>\alpha^*$}
        \label{cont2}
	\end{subfigure}
        \begin{subfigure}{.55\textwidth}
        \captionsetup{format=hang}
        \includegraphics[width=\textwidth]{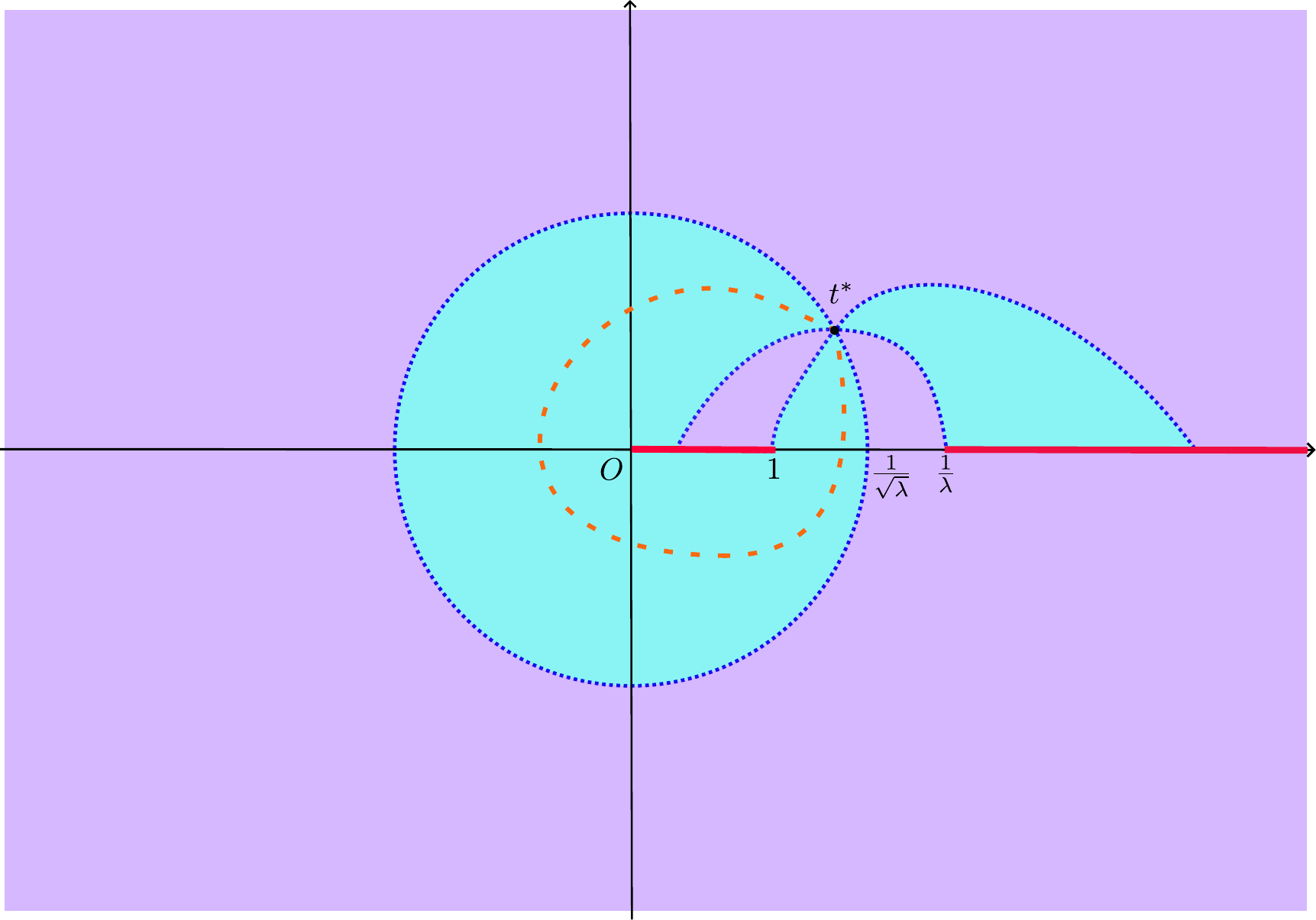}
		\caption{The integration contour for $\alpha=\alpha^*$}
        \label{cont3}
	\end{subfigure}
        \captionsetup{format=hang}
	\caption{A sketch of integration contours, black dots are saddle points, red lines 
        are branch cuts of the function $G(t)$, blue dotted lines are level curves with constant $\text{Re}[G(t)]$, orange dashed lines are integration contours. Areas with $\text{Re}[G(t)]$ grater/smaller than that at the saddle points passed by the integration contour are filled with purple/blue. Note that level curves either close or end on branch cuts, and only intersect at saddle points.}
        \label{intcontour}
\end{figure}

In this appendix we explain in detail the saddle point analysis of the complex integral~\ref{integral}. The function $G(t)$ defined in~\ref{exponent} has two saddle points as given by Equation~\ref{saddle}. Let $\alpha^*=\frac{2\sqrt\lambda}{1-\lambda}$, when $\alpha<\alpha^*$ the two saddles sit on the semicircle centered at $0$ with radius $\frac{1}{\sqrt\lambda}$, and when $\alpha>\alpha^*$ they sit on the semicircle centered at $\frac{\lambda+1}{2\lambda}$, with a radius $\frac{1-\lambda}{2\lambda}$, as illustrated in Figure~\ref{figure for saddles}. 
The integration contour in Equation~\ref{integral} is a closed curve which encircles the segment $[0,1]$ and does not touch the branch cut of $G(t)$. To implement the steepest descent method, we deform the contour to pass through the saddle points $t_s$, while ensuring that $\text{Re}[G(t)]$ reaches its global maxima along the contour at the saddle points. That is, the deformed contour has to lie entirely within the \emph{valleys} where $\text{Re}[G(t)]<\text{Re}[G(t_s)]$ and cannot trespass into the \emph{hills} where $\text{Re}[G(t)]>\text{Re}[G(t_s)]$. Furthermore, we require the contour to be the \emph{steepest descent contour} in the neighbourhood of the saddle points, as it can be shown that $\text{Im}[G(t)]$ is a constant along this contour, which eliminates uncontrolled oscillation. Now we can use the Laplace method around each saddle point to approximate the integral. When $\alpha\neq\alpha^*$, both $t_1$ and $t_2$ are second order saddles, which means that $G''(t_{1,2})\neq 0$. The neighbourhood of a second order saddle point contains two hills and two valleys in an alternate way, separated by level curves on which $\text{Re}[G(t)]=\text{Re}[G(t_s)]$. In order to understand how to close the contour without trespassing into the hills we need a careful analysis of the global property of $\text{Re}[G(t)]$ on the complex plane. For $\alpha<\alpha^*$, we have $\text{Re}[G(t_1)]=\text{Re}[G(t_2)]=\alpha\ln\frac{1}{\sqrt\lambda}$. In this case hills and valleys of $\text{Re}[G(t)]$ are illustrated in Figure~\ref{cont1},  
and we have to pick up both saddle points in order to close the integration contour without trespassing into the hills. Standard steepest descent approximation then gives Equation~\ref{intasymp1}, where $\phi_{1,2}$ are determined by the directions of steepest descent paths at $t_{1,2}$ (we have defined the positive direction to be counter-clockwise)
\begin{align}
    \phi_1&=\frac{3\pi-\arg[G''(t_1)]}{2}\\
    \phi_2&=\frac{\pi-\arg[G''(t_2)]}{2}
\end{align}
Similarly, for $\alpha>\alpha^*$, hills and valleys of $\text{Re}[G(t)]$ are shown in Figure~\ref{cont2}. In this case we have to pick up $t_2$ only. Therefore the steepest descent method gives Equation~\ref{intasymp2}, where the angle $\phi_2$ is given by
\begin{equation}
    \phi_2=\frac{\pi-\arg[G''(t_2)]}{2}
\end{equation}

\section{Evaluation of the Integrals}
\label{append:integral}
In this appendix we provide a detailed evaluation of integrals in Equation~\ref{threeterms}. It is easy to see that to the lowest order $G''(t_{1,2})\sim \sqrt\delta$ when $\alpha=\alpha^*+\delta$ (see below), hence the width of the second region is roughly $\delta\propto\frac{1}{s^2}$, and within this region we can estimate $f(c,s)$ using its value at $\alpha=\alpha^*$. As we will demonstrate later, saddle point analysis gives $I(s\alpha^*,s)\sim s^{-\frac{1}{3}}$ and thus the contribution $R_2(s)$ from $(\alpha^*-\delta,\alpha^*+\delta)$ goes as $s^{-\frac{2}{3}}$, which vanishes for large $s$. Next we calculate the contribution from the first and the third region. Since $G''(t_{1,2})\sim \sqrt\delta$ around $\alpha^*$, these two integrals converge as we take the limit $s\rightarrow \infty$ (or equivalently $\delta\rightarrow 0$). The contribution from the first region is given by plugging Equation~\ref{asymp1} into the integral~\ref{integral ramp}, the oscillating term averages to $0$ for large $s$, thus we have
\begin{equation}
    R_1(s)=\frac{\lambda s}{4\pi^2}\int_0^{\alpha^*} d\alpha\; \alpha
    \bigg\{
    \frac{2\pi}{|G''(t_1)|}+\frac{2\pi}{|G''(t_2)|}
    \bigg\}
\end{equation}
This term is linear in $s$ since $t_{1,2}$ are solely functions of $\alpha$ and independent of $s$. Finally, we calculate the contribution from the third region, plugging Equation~\ref{asymp2} into~\ref{integral ramp} we have
\begin{equation}
    R_3(s)=\frac{\lambda s}{4\pi^2}\int_{\alpha^*}^\infty d\alpha\; \frac{2\pi\alpha}{|G''(t_2)|}
    e^{s\{\alpha\ln \lambda+2\text{Re}[G(t_2)]\}}
\end{equation}
where the exponential decays to $0$ as $\alpha$ increases. Thus $R_3(s)$ is subleading for large $s$.\footnote{We can see this without referring to the details of the function $\alpha\ln \lambda+2\text{Re}[G(t_2)]$. In fact, it is enough to ensure that the function appearing on the exponential decreases monotonically for large $s$. This is because such a function reaches its maximum in $[\alpha^*,\infty)$ either at $\alpha^*$ or at some $\alpha_0>\alpha^*$. Together with the fact that $G''(t_2)\sim\sqrt\delta$ around $\alpha^*$, we can easily find that in both cases the integral scales as $s^{\frac{1}{2}}$ and is therefore subleading.} Collecting the results above gives the result Equation~\ref{linearramp}. \\ \\
Now we only need to verify that $R_2(s)\sim s^{-\frac{2}{3}}$. This can be done by estimating the value of $f(c,s)$ near the critical point $\alpha^*$. When $\alpha=\alpha^*$ we have $t_1=t_2=t^*$ and $G''(t^*)=0$. However, we have $G^{(3)}(t^*)\neq 0$ at this point in general, and thus $t^*$ is a third order saddle point. In this case the steepest descent method is still applicable, with the leading order contribution determined by $G^{(3)}(t^*)$. The neighbourhood of a third order saddle point contains three hills and three valleys in an alternate way, separated by level curves of constant $\text{Re}[G(t)]$. Level curves with $\text{Re}[G(t)]=\text{Re}[G(t^*)]$ and the integration contour we use in this case are shown in Figure~\ref{cont3}. We can parametrize the integration contour by the curve length $x$, with $x=0$ at $t^*$. Note that the contour is not smooth (but still continuous) at $t^*$, thus we need to consider the contribution from $x>0$ and $x<0$ separately. For $x>0$ we have
\begin{equation}
    I_+\sim e^{sG(t^*)+i\phi_+}\int ds\; \exp(-\frac{sG^{(3)}(t^*)}{3!}x^3)=\frac{1}{3}\Gamma(\frac{1}{3})\bigg(\frac{3!}{sG^{(3)}(t^*)}\bigg)^{\frac{1}{3}}e^{sG(t^*)+i\phi_+}
\end{equation}
And similarly for $x<0$ we have
\begin{equation}
    I_-\sim e^{sG(t^*)+i\phi_-}\int ds\; \exp(-\frac{sG^{(3)}(t^*)}{3!}x^3)=\frac{1}{3}\Gamma(\frac{1}{3})\bigg(\frac{3!}{sG^{(3)}(t^*)}\bigg)^{\frac{1}{3}}e^{sG(t^*)+i\phi_-}
\end{equation}
Here the angles $\phi_{\pm}$ are again determined by the direction of the steepest descent paths:
\begin{align}
    \phi_+&=\frac{3\pi-\arg[G^{(3)}(t^*)]}{3}\\
    \phi_-&=\frac{5\pi-\arg[G^{(3)}(t^*)]}{3}
\end{align}
Thus we have proved that $I(c,s)\propto s^{-\frac{1}{3}}$, or $f(c,s)\sim s^{\frac{1}{3}}$ at $\alpha=\alpha^*$, as expected. \\ \\
Finally we verify the behaviour of $G''(t_{1,2})$ near the critical point $\alpha^*$. Note that $G''(t_{1,2})$ depends on $\alpha$ in two ways: explicitly through the term containing $\alpha$ in $G(t)$, and implicitly through $t_{1,2}$ which are functions of $\alpha$ only (as $\lambda$ is fixed). Since $G''(t^*)=0$, we can expect that for $\alpha=\alpha^*+\Delta\alpha$, it takes the following form
\begin{equation}\label{second order derivative expansion}
    G''(t_{1,2})\sim \frac{\partial G''(t_{1,2})}{\partial \alpha}\Delta\alpha+\frac{\partial G''(t_{1,2})}{\partial t_{1,2}}\Delta t_{1,2}
\end{equation}
where the partial derivatives are evaluated at $\alpha=\alpha^*$ and are generally non-zero. The first term is always of order $\Delta \alpha$, so we focus on the second term. In Equation~\ref{saddle}, when $\alpha=\alpha^*$, the term in the square root is $0$. Therefore, for $\alpha=\alpha^*+\Delta \alpha$ where $\Delta\alpha$ is a small number, we have $\Delta t_{1,2}\propto \sqrt {\Delta\alpha}$. Equation~\ref{second order derivative expansion} then tells that the leading order contribution to $G''(t_{1,2})$ around $\alpha^*$ is of order $\sqrt{\Delta\alpha}$, that is $G''(t_{1,2})\sim \sqrt{\Delta\alpha}$ for $\alpha=\alpha^*+\Delta\alpha$, which we have used above.

\section{Review of the Equilibrium Approach}
\label{append:equi}
In this appendix we review the details of the equilibrium approach to the single and double traces Equation~\ref{latetimetrace} and~\ref{latetimedoubletrace} discussed in~\cite{liu2021entanglement}. A useful way to think of the trace $\text{tr}_A(\text{tr}_B U\ket{\Psi_0}\bra{\Psi_0}U^\dagger)^n$ is to view it as the time evolution of $2n$ copies of the same system under $U(t)$, with some fixed boundary conditions. To be accurate, we need $n$ copies of the original system and another $n$ copies of its conjugate. If we take the basis of the original Hilbert space to be $\ket{i}$ then we pick the basis for the conjugate space to be $\ket{\bar i}=T\ket{i}$, where $T$ is some anti-unitary operator (such as the CPT operator). The evolution of the conjugate states are implied by $U^\dagger(t)$ and their inner products satisfy $\braket{\bar i|\bar j}=\braket{j|i}^*$. Before proceeding we introduce some notations which proves to be useful later. For the Hilbert space $(\mathcal H\otimes\bar{\mathcal H})^n$, we write its basis vector as
\begin{equation}
    \ket{\{i\}}\otimes\ket{\{\bar i'\}}\equiv \ket{i_1}\otimes\ket{\bar i'_1}\ldots\ket{i_n}\otimes\ket{\bar i'_n}
\end{equation}
here $\{i\}$ denotes a sequence of $i$'s. Now for a permutation $\sigma\in S_n$ we introduce the following state
\begin{equation}
    \ket{\sigma}=\sum_{\{i\}}\ket{\{i\}}\otimes\ket{\sigma\{\bar i\}}
\end{equation}
Here $\sigma\{i\}$ is the sequence obtained by acting $\sigma$ on the sequence $\{i\}$, and the sum is over all sequences. This state is not normalized, and satisfies 
\begin{equation}
    \braket{i_1\bar i_1'i_2\bar i_2'\ldots i_n\bar i_n'|\sigma}=\delta_{i_1 i'_{\sigma(1)}}
    \delta_{i_2 i'_{\sigma(2)}}\ldots\delta_{i_n i'_{\sigma(n)}}
\end{equation}
Furthermore for an operator $O$ we define the following state
\begin{equation}
    \ket{O,\sigma}=\sum_{\{i\}}O^{\otimes n}\ket{\{i\}}\otimes\ket{\sigma\{\bar i\}}
\end{equation}
note that here we only act $O$'s on the $\mathcal H^{\otimes n}$ part but not on the $\bar{\mathcal H}^{\otimes n}$ part. This notation is useful because for an arbitrary state $\ket{\Psi_0}$ we can always define a density operator $\rho_0=\ket{\Psi_0}\bra{\Psi_0}$, which satisfies
\begin{equation}
    (\ket{\Psi_0}\otimes\ket{\bar\Psi_0})^n=\ket{\rho_0,e}
\end{equation}
where $e$ is the unit element of $S_n$. Another frequently used property is that, for any two states $\ket{O_1,\tau}$ and $\ket{O_2,\sigma}$ we have 
\begin{equation}
    \braket{O_1,\tau|O_2,\sigma}=\text{tr}(O_1^\dagger O_2)^{n_1}
    \text{tr}(O_1^\dagger O_2)^{n_2}\ldots \text{tr}(O_1^\dagger O_2)^{n_k}
\end{equation}
where $k$ is the number of orbits in $\sigma \tau^{-1}$ and $n_k$'s are the length of each orbit respectively. Finally in order to deal with bipartite systems we have to introduce basis independently for A and B subsystems, thus we need to generalize our basis defined above to the following form
\begin{equation}
    \ket{\{i_a\},\{j_b\}}\otimes\ket{\{\bar i'_a\},\{\bar j'_b\}}\equiv \ket{i_{1a}}\otimes\ket{j_{1b}}\otimes\ket{\bar i'_{1a}}\otimes\ket{\bar j'_{1b}}\ldots\ket{i_{na}}\otimes\ket{j_{nb}}\otimes\ket{\bar i'_{na}}\otimes\ket{\bar j'_{nb}}
\end{equation}
and we also define the \emph{subsystem permutations} which act only on subsystem A or B, such a permutation can be written as $\sigma\otimes \tau$, for which we define the following state
\begin{equation}
    \ket{\sigma\otimes \tau}=\sum_{\{i\},\{j\}}\ket{\{i_a\},\{j_b\}}\otimes\ket{\sigma\{\bar i_a\},\tau\{\bar j_b\}}
\end{equation}
With this setup we can immediately write
\begin{equation}\label{renyiamplitude}
    S_A^{(n)}=\braket{\eta_A\otimes e_B|(U\otimes U^\dagger)^{\otimes n}|\rho_0,e}
\end{equation}
and 
\begin{equation}\label{correlatoramplitude}
    \text{tr}\rho_A^n\;\text{tr}\rho_A^m=\braket{\gamma_{0A}\otimes e_B|(U\otimes U^\dagger)^{\otimes n+m}|\rho_0,e}
\end{equation}
as before we have $\eta=(123\ldots n)$ and $\gamma_0=(123\ldots n)(n+1,n+2\ldots n+m)$. In order to put these amplitudes into a unified form we define, for an arbitrary subsystem permutation $\mu$
\begin{equation}
    Z_n(\mu)=\braket{\mu|(U\otimes U^\dagger)^{\otimes n}|\rho_0,e}
\end{equation}
Next we define the \emph{effective identity operator} $I_\alpha$ by 
\begin{equation}
    \rho_{eq}=\frac{I_\alpha}{Z(\alpha)}
\end{equation}
where $\rho_{eq}$ is the equilibrium density matrix defined above and $Z(\alpha)$ is the respective partition function, $\alpha$ is an index which denotes which ensemble we are using. In the case of Haar ensemble $I_\alpha$ is just the identity operator $\mathbb I$. Furthermore we define
\begin{equation}
    Z_n(\alpha)=\text{tr}I_\alpha^n
\end{equation}
with these notations we can write (from now on we will neglect the index $\alpha$ for simplicity)
\begin{equation}
    \braket{I_\alpha,\tau|I_\alpha,\sigma}=Z_{2n_1}Z_{2n_2}\ldots Z_{2n_k}
\end{equation}
where $k$ is again the number of orbits in $\sigma \tau^{-1}$. A specific case is that
\begin{equation}
    \braket{I_\alpha,\sigma|I_\alpha,\sigma}=Z_2^n
\end{equation}
Now we can introduce a metric on the subspace expanded by states of the form $\ket{I_\alpha,\sigma}$, note that these states are not orthogonal to each other:
\begin{equation}\label{metric}
    g_{\tau\sigma}=\frac{\braket{I_\alpha,\tau|I_\alpha,\sigma}}{(\braket{I_\alpha,\tau|I_\alpha,\tau})^{\frac{1}{2}}(\braket{I_\alpha,\sigma|I_\alpha,\sigma})^{\frac{1}{2}}}=\frac{Z_{2n_1}Z_{2n_2}\ldots Z_{2n_k}}{Z_2^n}
\end{equation}
the projector onto this subspace is therefore
\begin{equation}
    P_\alpha=\frac{1}{Z_2^n}\sum_{\tau,\sigma} g^{\tau\sigma}\ket{I_\alpha,\tau}\bra{I_\alpha,\sigma}
\end{equation}
where $g^{\tau\sigma}$ is the inverse of $g_{\tau\sigma}$. Note that, since $I_\alpha$ is supposed to be invariant under time evolution $U(t)$, $P_\alpha$ is invariant under $(U\otimes U^\dagger)^{\otimes n}$. The key proposal of~\cite{liu2021entanglement} is that, after the equilibrium time, we can effectively replace $(U\otimes U^\dagger)^{\otimes n}$ by $P_\alpha$. This result is exact in the Haar random case which we are mainly interested in as it can be proved that~\cite{zhou2020entanglement,gu2013moments}
\begin{equation}
    \braket{(U\otimes U^\dagger)^{\otimes n}}_{Haar}=P_{\mathbb I}
\end{equation}
Now we make this replacement in Equation~\ref{renyiamplitude} and~\ref{correlatoramplitude}. Which gives the general expression
\begin{equation}
    Z_n(\mu)=\frac{1}{Z_2^n}\sum_{\sigma,\tau}g^{\tau\sigma}\braket{\mu|I_\alpha,\tau}\braket{I_\alpha,\sigma|\rho_0,e}
\end{equation}
Of course, here we are summing over permutations in the group $S_n$. On the other hand, the assumption that averaging over $(U\otimes U^\dagger)^{\otimes n}$ is equivalent to replacing it by the projector $P_\alpha$ gives, in the special case $n=1$
\begin{equation}
    \text{tr}(I_\alpha \rho_0)=\frac{Z_2}{Z_1}
\end{equation}
this comes from the fact that $\text{tr}_A\text{tr}_B( U\ket{\Psi_0}\bra{\Psi_0}U^\dagger)=1$, which survives the averaging process. Furthermore, since $\rho_0$ is a pure state density matrix we have 
\begin{equation}
    \text{tr}(\rho_0 I_\alpha)^n=\braket{\Psi_0|I_\alpha|\Psi_0}^n=[\text{tr}(\rho_0 I_\alpha)]^n=\frac{Z_2^n}{Z_1^n}
\end{equation}
Using these identities we can rewrite $Z_n(\mu)$ as
\begin{equation}
    Z_n(\mu)=\frac{1}{Z_1^n}\sum_{\sigma,\tau}g^{\tau\sigma}\braket{\mu|I_\alpha,\tau}=\frac{a}{Z_1^n}\sum_{\tau}\braket{\mu|I_\alpha,\tau}
\end{equation}
where $a=\sum_\sigma g^{\tau\sigma}$. As we always assume the dimension of the Hilbert space of the system is a large number, we expect each $Z_n$ to contribute a large factor of the same order as $Z_1$, which can be seen as the \emph{effective dimension} of the subspace defined by the projector $P_\alpha$. Thus to the leading order Equation~\ref{metric} gives 
\begin{equation}
    g^{\tau\sigma}=\delta^{\tau\sigma}+O(\frac{1}{Z_1})
\end{equation}
therefore our expression for $Z_n(\mu)$ simplifies to
\begin{equation}
    Z_n(\mu)\sim\frac{1}{Z_1^n}\sum_{\tau}\braket{\mu|I_\alpha,\tau}
\end{equation}
in the case of double trace correlator it gives
\begin{equation}
    \braket{\text{tr}\rho_A^n\;\text{tr}\rho_A^m}=\frac{1}{Z_1^{n+m}}\sum_{\tau}\braket{\gamma_{0A}\otimes e_B|I_\alpha,\tau}
\end{equation}
Here the braket means that we have taken the equilibrium approximation. After expanding the expression above and some tedious algebra, we have
\begin{equation}\label{double trace equilibrium approximation}
    \braket{\text{tr}\rho_A^n\;\text{tr}\rho_A^m}_c=\frac{1}{Z_1^{m+n}}\sum_{\substack{\{i\},\{j\}\\ \tau\;\text{connected}}}
    \braket{i_{\tilde\tau(1)}j_{\tau(1)}|I_\alpha|i_1j_1}\ldots \braket{i_{\tilde\tau(n+m)}j_{\tau(n+m)}|I_\alpha|i_{n+m}j_{n+m}}
\end{equation}

\section{Derivation of Equation~\ref{microcanonical double trace}}
\label{append:micro}
Here we derive Equation~\ref{microcanonical double trace} from Equation~\ref{double trace equilibrium approximation}. Plugging the projector Equation~\ref{micro projector} into Equation~\ref{double trace equilibrium approximation}, we get a $n+m$ fold sum over energies $E_A^i$, where $i=1,2\ldots,n+m$. Therefore we get the following expression for the sum in Equation~\ref{double trace equilibrium approximation}
\begin{equation}\label{micro traces}
    \sum_{E_A^i} \prod_{\tilde\sigma,\sigma} \text{tr}\big[\prod_{j\in\tilde\sigma} P_A(E_A^{j})\big] \text{tr}\big[\prod_{k\in\sigma} P_B(E-E_A^{k})\big]
\end{equation}
Here $\tilde\sigma,\sigma$ are orbits of the permutation $\tilde \tau,\tau$. Since $P_A$ are projectors we have  
\begin{equation}
    P_A(E_A)P_A(E_A')=\begin{cases}
        P_A(E_A)\;\;&(E_A=E_A')\\
        0\;\;&(E_A\neq E_A')
    \end{cases}
\end{equation}
and the same for $P_B$. From Equation~\ref{micro traces} we can see that for each pair of $E_A^i$ and $E_A^j$, if there is a sequence $\mu_k$ ($k=1,2\ldots$) where each $\mu_k$ is either $\tau$ or $\tilde\tau$, such that $\mu_1\circ\mu_2\circ\mu_3\ldots (i)=j$, then we have a constraint that $E_A^i=E_A^j$. In fact it follows that we only have one independent $E_A$ in the sum. If this is not the case, then there must exist a subset $S$ of the $n+m$ elements which is invariant under both $\tau$ and $\tilde\tau$, however this implies that $S$ is invariant under $\gamma_0$, which is impossible unless $S$ is empty or contains all the elements (since the permutation is connected $S$ cannot be the set of all elements on one of the circles). Thus we have proved Equation~\ref{microcanonical double trace}.


\bibliographystyle{JHEP}
\bibliography{biblio.bib}


\end{document}